\documentclass[a4paper,12pt]{article}
\pdfoutput=1
\usepackage{jheppub}
\usepackage{amssymb, amsmath}
\usepackage{graphicx}
\usepackage{hyperref}

\usepackage{tikz}
\usetikzlibrary{decorations.markings,
  decorations.pathreplacing,arrows}
\usetikzlibrary{shapes.geometric}
\usetikzlibrary{decorations.pathmorphing,decorations.text}

\begin{document}
\begin{flushright}
INR--TH--2020--023
\end{flushright}
\vspace{-1cm}

\title{Dilaton gravity with a boundary: from unitarity to black hole
  evaporation}
\author[a, b]{Maxim Fitkevich,}
\author[a, c]{Dmitry Levkov,}
\author[b, c, d]{and Yegor Zenkevich}
\affiliation[a]{\footnotesize Institute for Nuclear Research of the Russian Academy
  of Sciences, Moscow 117312,
  Russia}
\affiliation[b]{\footnotesize Moscow Institute of Physics and Technology,
  Dolgoprudny 141700, Moscow Region, Russia}
\affiliation[c]{\footnotesize Institute for Theoretical and Mathematical Physics, 
  MSU, Moscow 119991, Russia}
\affiliation[d]{\footnotesize SISSA, Trieste  34136, Italy; 
  INFN, Sezione di Trieste, Italy; IGAP, Trieste 34151, Italy\\
  Institute for Theoretical and Experimental Physics, Moscow 117218, Russia}
\emailAdd{fitkevich@phystech.edu}
\emailAdd{levkov@ms2.inr.ac.ru}
\emailAdd{yegor.zenkevich@gmail.com}

\abstract{We point out that two-dimensional Russo-Susskind-Thorlacius
  (RST)  mo\-del for evaporating black holes is locally equivalent~---
  at the full quantum level~--- to flat-space Jackiw-Teitel\-boim (JT) gravity that
  was recently shown to be unitary. Globally, the two models differ by
  a reflective spacetime boundary added in the RST model. Treating the
  boundary as a local and covariant deformation of quantum JT theory,
  we develop sensible semiclassical description of evaporating RST
  black holes. Nevertheless, our semiclassical solutions
  fail to resolve the information recovery problem, and they do not
  indicate formation of remnants. This means that either the
  standard semiclassical method incorrectly describes the evaporation process
  or the RST boundary makes the flat-space JT model fundamentally inconsistent.}

\maketitle

%%%%%%%%%%%%%%%%%%%%%%%%%%%%%%%%
\section{Introduction}
\label{sec:intro}
\begin{figure}[t]
  %% \unitlength=1mm
  %% \begin{picture}(100,10)
  %%   \put(0,0){\line(0,-1){95}}
  %%   \put(155,0){\line(0,-1){150}}
  %%   \put(0,-92.5){\line(1,0){150}}
  %%   \put(0,-5){\line(1,0){150}}
  %% \end{picture}
  
  \centerline{
    \hspace{-1mm}\begin{minipage}{6cm}
      \vspace{-5.8mm}
      \includegraphics{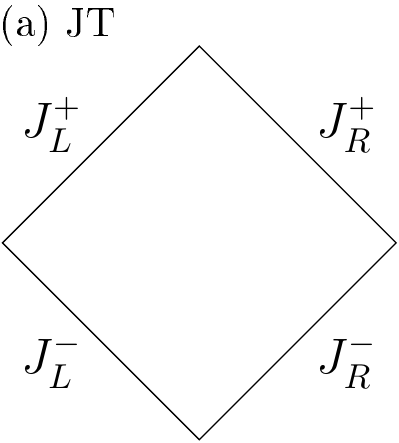}\\[1.1ex]
      \includegraphics{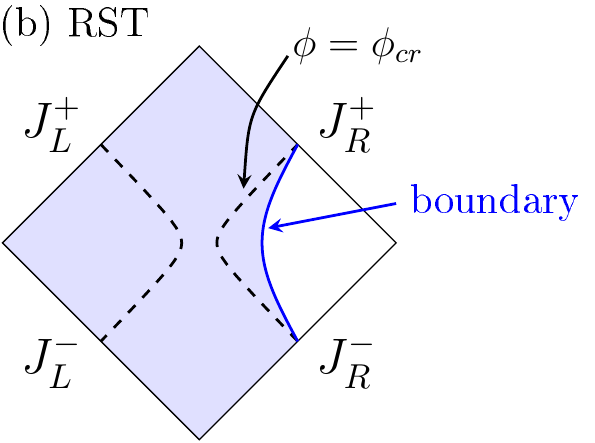}        
    \end{minipage}
  \hspace{3mm}
  \begin{minipage}{7.8cm}
    \includegraphics{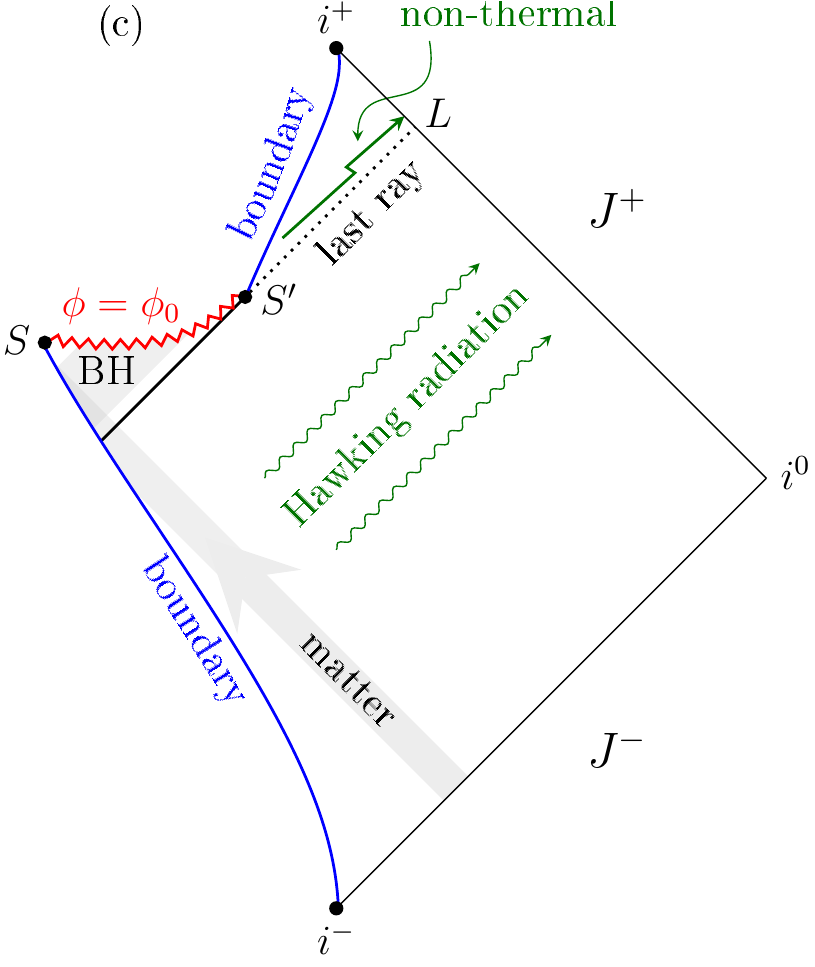}
  \end{minipage}
  }
  \begin{picture}(100,1)
  \end{picture}
  \caption{(a), (b) Causal structures of flat-space JT model and RST
    model. (c)~Penrose diagram for the evaporating RST black
    hole.\label{fig:intro}}
\end{figure}

Recently the simplest theory of two-dimensional dilaton gravity~---
flat-space Jackiw-Teitelboim (JT) model~\cite{Jackiw:1984je,
  Teitelboim:1983ux, Cangemi:1992bj}~--- was quantized and its nontrivial, explicitly
unitary ${\cal S}$-matrix was obtained~\cite{Dubovsky:2017cnj,
  Dubovsky:2018bmo}, see also~\cite{Saad:2019lba, Stanford:2019vob}. This
model displays so many features of full  multidimensional gravity that
one can hastily anticipate  its 
application to the long-standing puzzles of black hole physics like
information paradox~\cite{Hawking:1974sw, Hawking:1976ra,
  Maldacena:2001kr, Penington:2019npb, Almheiri:2019psf,
  Penington:2019kki, Gautason:2020tmk}, firewall proposal~\cite{Almheiri:2012rt}
(cf.~\cite{Penington:2019npb}), or
non-conservation of global charges~\cite{Zeldovich1976, Coleman:1993zz, 
  Stojkovic:2005zq}. However, the JT metric is flat on field
equations, and all classical solutions in this theory are 
causally equivalent to empty two-dimensional spacetime, see 
Fig.~\ref{fig:intro}a. This  precludes formation of authentic black
holes with horizons and singularities.  Introducing $N$ matter fields
$\hat{f}_i(x)$, one can unitarily transfer information between the
past and future JT  infinities, but learn nothing about the black hole 
physics.

In this paper we consider the celebrated Russo-Susskind-Thorlacius
(RST) mo\-del~\cite{Russo:1992ax} which is specifically designed to
describe evaporating black holes in two dimensions. We point out that
this model is locally equivalent, at the full quantum level, to the
unitary flat-space JT gravity. To establish the equivalence, one promotes
the one-loop RST model to a full quantum theory: one
adds RST counter-term~\cite{Russo:1992ax} and $N$ matter
fields $f_i(x)$ to the action of dilaton
gravity~\cite{Callan:1992rs}, and then quantizes the resulting
  theory in a consistent way suggested by
Strominger~\cite{Strominger:1992zf}. After that the path integrals of
JT and RST models are related by the local Weyl
transformation\footnote{Equivalence at the classical level was
  previously established in~\cite{Cangemi:1992bj,  Dubovsky}. 
  Flatness of Weyl-transformed one-loop geometry was pointed out
  in~\cite{Strominger:1992zf}. Also, it is known that
    redifinition~\eqref{eq:5} of the dilaton field is
    required~\cite{Russo:1992ax} to solve the RST model at the 
    semiclassical level.},
\begin{equation}
  \label{eq:5}
  \hat{g}_{\mu\nu} = \mathrm{e}^{-2\phi} \, g_{\mu\nu} \;, \qquad\qquad
  \hat{\phi} = \mathrm{e}^{-2\phi} + \frac{N \phi}{48 \pi}\;, \qquad\qquad
  \hat{f}_i = f_i\;,
\end{equation}
where $g_{\mu\nu}(x)$, $\phi(x)$ and $f_i(x)$ are the RST metric,
dilaton and matter fields, while $\hat{g}_{\mu\nu}(x)$,
$\hat{\phi}(x)$ and $\hat{f}_i(x)$ are the fields in the flat-space JT
model. Note that the change of variables (\ref{eq:5}) causes
anomalous transformation of the functional measure which is important
for the equivalence.  In what follows we denote $Q^2 \equiv N/(48 \pi)$.

Unfortunately, the transformation~(\ref{eq:5}) is not defined
globally. Indeed, it is non-invertible near the critical value of
the dilaton field ${\phi_{cr} = -\frac12\log (Q^2/2)}$ corresponding to
${d\hat{\phi}/d\phi=0}$. This is a problem because all finite-energy
JT/RST solutions, e.g.\ the linear dilaton vacuum, have
position-dependent dilaton reaching $\phi_{cr}$ at some critical curve
(dashed line in Fig.~\ref{fig:intro}b). The transformation from the
healthy JT theory to the RST variables is singular at this
curve. Early works observed~\cite{Banks:1992ba, Russo:1992ht, Russo:1992ax,
  Thorlacius:1994ip} that the semiclassical solutions
describing evaporating RST black holes develop curvature singularities
at ${\phi=\phi_{cr}}$, and this impedes quantization of the
theory~\cite{deAlwis:1992emy}.  We show that these singularities
disappear in full quantum theory after transforming to the JT terms.

It would be natural to equate the RST and flat-space JT models discarding
singular pa\-ra\-me\-tri\-za\-tion by the RST fields. But that would
degrade the spectacular RST black holes to fictitious
objects\footnote{These objects are different from the JT analogs of
  black holes introduced in~\cite{Brown:1986nm}.} hiding
information under the fake singularities of the RST fields,
cf.~\cite{Almheiri:2013wka}.  

In this paper we explore another possibility suggested in the original
works~\cite{Russo:1992ax, Russo:1992yh, Chung:1993rf}. Namely, we
deform the RST model: we introduce a reflective spacetime
boundary along the line of constant dilaton $\phi(x) = \phi_0<
\phi_{cr}$ and restrict all fields in the path integral to the
submanifold $\phi \leq \phi_0$ (white region in
Fig.~\ref{fig:intro}b). This excludes the Weyl singularity from the
physical domain and makes the model causally similar to the
spherically-symmetric multidimensional gravity.

Unlike in the earlier studies, we have a solid tool for selecting
sensible boundary conditions at $\phi = \phi_0$. Indeed, we 
first add the boundary to the healthy flat-space JT theory, making it a
local, covariant, self-consistent, and weakly coupled deformation,
and then transform  to the RST terms. Moreover, since 
(\ref{eq:5}) is  valid at the quantum level, we compute one-loop
effective   action~\cite{Polyakov:1981rd} with correct boundary
terms~\cite{Polchinski:1998rq} in the quantum JT theory and then perform 
Weyl transformation. This gives one-loop RST boundary conditions 
which automatically satisfy all self-consistency
criteria~\cite{Strominger:1994xi}. Our reflection laws at $\phi =
\phi_0$ are 
similar to those in~\cite{Chung:1993rf, Das:1994yc} but differ from
the laws in~\cite{Russo:1992yh,  Verlinde:1993sg,
  Bose:1995bk, Bose:1996pi}.

Once the self-consistent boundary conditions are found, we re-inspect
information loss problem in the RST model. To this end we
study, both analytically~\cite{Fitkevich:2017izc} and numerically,
the semiclassical solutions extremizing the one-loop effective
action. A typical high-energy solution is shown in
Fig.~\ref{fig:intro}c. It still displays some of the
undesirable features observed in the earlier
studies~\cite{Strominger:1994tn}. In particular, $\phi(x)$ equals
$\phi_0$ at three distinct lines: the timelike boundaries
  $i^{-}S$ and $S'i^+$, and a spacelike curve $SS'$. We cannot
impose reflective boundary condition at $SS'$, as it would imply
strong violation of causality, but we still have to trim the spacetime
along this line.  Thus, $SS'$ is an analog of black hole
singularity in the RST model. The incoming matter irreversibly disappears behind this
line, cf.~\cite{Russo:1992ax}.

We find that the spacetime of evaporating black hole can be
continued into the future beyond the last ray $S'L$. Indeed, the
``singularity'' line $\phi = \phi_0$ generically becomes timelike
after some point  $S'$~--- the endpoint of
  evaporation. Imposing the boundary conditions in 
    that region, we obtain the branch $S'i^+$ of
reflective boundary in Fig.~\ref{fig:intro}c. This makes the
spacetime flat in the asymptotic future $i^+$. The price to pay is
the jump of second metric derivatives at the last
ray~$S'L$ leading to a small $\delta$-burst of negative
energy along this ray (``thunderpop'' in~\cite{Russo:1992ax,
  Strominger:1994tn}). The latter inconsistency, however, appears
  due to sharp change of boundary conditions across the point
$S'$. Once ``smearing'' near $S'$ is allowed, the ``thunderpop''
becomes smooth.

Despite our effort to construct a sensible model, the semiclassical
solution in Fig.~\ref{fig:intro}c still indicates an apparent loss of
quantum coherence by the evaporating black hole. The effect is
caused by the singularity $SS'$ irreversibly ``eating'' the pure
quantum state of the incoming matter. We support this
intuition by computing the entanglement entropy~\cite{Holzhey:1994we,
  Bianchi:2014bma, Good:2016atu} of the final state at the future
  null infinity $i^0 i^+$. We demonstrate that large entropy of
Hawking radiation at $i^0 L$ cannot be compensated by
the non-thermal radiation at $Li^+$. Thus, initial pure state of
matter transforms into a density matrix with nonzero entanglement, and
unitarity is broken. 

The same mechanism that ruins quantum coherence implies apparent 
non-con\-ser\-va\-tion of global charges by the evaporating black
holes~\cite{Zeldovich1976, Coleman:1993zz, Stojkovic:2005zq}. Indeed, 
our model possesses a global shift charge, and we
explicitly demonstrate that it disappears behind the black
hole singularity $SS'$.

To summarize, in this paper we try to make sense of the evaporating RST
black holes by appealing to the healthy flat-space JT theory deformed by a
boundary. Nevertheless, we observe apparent information loss
inside these objects. Possible reasons for this failure will
  be discussed in Sec.~\ref{sec:discussion}.

%%%%%%%%%%%%%%%%%%%%%%%%%%%%%%%%
\section{From RST to JT}
\label{sec:jt-to-rst}
\subsection{Weyl transformation}
\label{sec:equivalence-bulk}
Two-dimensional Russo-Susskind-Thorlacius (RST)
model~\cite{Russo:1992ax} describes interaction of $N$ matter fields
$f_j(x)$ with non-dynamical gravitational sector: metric
$g_{\mu\nu}(x)$ and dilaton $\phi(x)$. The action of the model 
\begin{equation}\label{eq:6}
S_{RST}=\int d^2x\,\sqrt{-g}\left[e^{-2\phi}\left(R+4(\nabla\phi)^2
+4\lambda^2\right)-\frac12\sum_{j=1}^N(\nabla f_j)^2-Q^2\; \phi R \right]\;,
\end{equation}
includes the classical part representing CGHS dilaton
gravity~\cite{Callan:1992rs} and a quantum
counter-term~\cite{Russo:1992ax}~--- the last term in the integrand~---
providing exact solvability at one-loop level. The parameter $Q^2
\equiv N/(48 \pi)$ is proportional to the number of scalar
fields, while $\lambda$ sets the energy scale of the model.

It will be important for us that the above model has a family of
classical black hole solutions extremizing the classical action
(\ref{eq:6}) without the counter-term, see~\cite{Callan:1992rs}. These
objects evaporate once quantum corrections are taken into
account. However, naive description of the evaporation reveals a
pathology~\cite{Strominger:1992zf} that has to be cured by
choosing a proper quantization procedure.
\begin{sloppy}

Usually, one quantizes (\ref{eq:6}) by fixing the
conformal gauge,
\begin{equation} 
  \label{eq:7}
  g_{\mu\nu} = \mathrm{e}^{2\rho(x)} \, \eta_{\mu\nu}\;,
\end{equation}
and introducing the standard Faddeev-Popov
ghosts~\cite{Polchinski:1998rq} $b_{\mu\nu}$ and $c^\mu$ with
$b_{\mu\nu} - b_{\nu\mu} = b^\mu_\mu = 0$ and action
${S_{gh}  =  -\int  d^2 x \sqrt{-g}\; b_{\mu\nu}\nabla^{\mu}
  c^\nu}$. This gives path integral
\begin{equation}
  \label{eq:8}
  Z_{RST}^{\mathrm{naive}} = \int [d\rho \, d\phi \, db \, dc\, df]_{\rho} \;
  \mathrm{e}^{iS_{RST}[\rho,\, \phi,\, f] + iS_{gh}[b,\, c]}\;,
\end{equation}
where all functional measures depend on the metric scale factor
$\rho(x)$ due to the Weyl anomaly. Now, we can evaluate
Eq.~(\ref{eq:8}) in the one-loop approximation. To this end we
integrate out quadratic fluctuations of all fields in some fixed
background~--- say, the black hole metric. We obtain
${Z_{RST}^{\mathrm{naive}} \approx \mathrm{e}^{iS_{RST} +
    iS_{\mathrm{1-loop}}}}$, where the correction is given by the
non-local Polyakov action~\cite{Polyakov:1981rd},
\begin{equation}
  \label{eq:10}
  S_{\mathrm{1-loop}} = -\frac{c}{96 \pi}\int d^2 x \,d^2x'\sqrt{gg'} \;
  R\, \Box^{-1}(x,x')R'\;.
\end{equation}
In Eq.~(\ref{eq:10}) we introduced the Green's function $\Box^{-1}$ of
the d'Alembertian and denoted the total central charge of all fields
by $c=N-24$. Extremizing the effective action~(\ref{eq:6}),
(\ref{eq:10}) with respect to the background metric and fields, one
obtains one-loop semiclassical equations, with solutions describing
evaporating black holes, see~\cite{Callan:1992rs, Ashtekar:2010hx,
  Ashtekar:2010qz} and~\cite{Russo:1992ax, Chung:1993rf, Bose:1995bk}.

 \end{sloppy}
 The problem is that the Hawking  flux from these objects is always
proportional~\cite{Callan:1992rs} to the factor in front of the one-loop
action~(\ref{eq:10})~--- the total central charge~$c=N-24$. The
latter, however, receives contributions from the entire
field content of the model: $N$ from matter fields, $+2$ from 
non-dynamical fields $\phi$ and $\rho$, and $-26$ from ghosts. This
means that at $N<24$ the black holes emit mainly ghosts and the total
energy flux is negative. At larger $N$ the total flux, though
positive, is not proportional to the number of dynamical fields $f_j$.  

One may wonder, why Hawking evaporation of unphysical Faddeev-Popov
ghosts is included in Eq.~(\ref{eq:10}). But this effect is
unavoidable! Indeed, in Heisenberg picture the ghost operators
$b_{\mu\nu}$ and $c^\mu$ satisfy causal equations in the black hole
background, just like the ordinary fields. As a consequence, their
positive- and negative-frequency components get mixed during
evolution between the horizon and asymptotic infinity. Then the same
Heisenberg vacuum that is ghost-free at the horizon, automatically
contains a flux of ghosts at infinity. This suggests that Eq.~(\ref{eq:8})
is physically inconsistent.

Strominger proposed~\cite{Strominger:1992zf} a simple way to avoid
this pathology. Namely, let us use the Weyl-transformed metric
$\hat{g}_{\mu\nu} = \mathrm{e}^{-2\phi(x)} \, g_{\mu\nu}$ with the scale
factor $\rho-\phi$ in the functional measures of all non-dynamical fields
$\rho$, $\phi$, $b$, $c$,
\begin{equation}
  \label{eq:12}
  Z_{RST} = \int [d\rho \, d\phi \, db \, dc]_{\rho -   \phi}\,
  [df]_{\rho} \; \mathrm{e}^{iS_{RST}[\rho,\, \phi,\, f] + iS_{gh}[b,\, c]}\;.
\end{equation}
Below we will see that any classical solution in the
model~(\ref{eq:6}) has flat $\hat{g}_{\mu\nu}$. Then the unphysical
fields are not emitted from the black 
holes. Indeed, loop contributions from these fields still produce the
Polyakov action (\ref{eq:10}), but with the curvature $\hat{R}$ of
$\hat{g}_{\mu\nu}$ replacing $R$. The effect of these contributions
vanishes at $\hat{R}=0$ implying that the Hawking flux
in Eq.~(\ref{eq:12}) is strictly 
proportional to the number $N$ of dynamical fields.

In what follows we use Eq.~(\ref{eq:12}) as a natural quantization
of the RST model and do not consider alternative prescriptions.

Let us now argue that the quantum RST model is locally equivalent to
the quantum flat-space JT theory~\cite{Jackiw:1984je,
  Teitelboim:1983ux, Cangemi:1992bj}. To this  
end we rewrite the path integral (\ref{eq:12}) in variables
(\ref{eq:5}) taking into account Weyl transformation law of
the functional measure~\cite{Polchinski:1998rq, Grumiller:2002nm},
\begin{equation}
  \label{eq:11}
  [df]_{\rho} = [df]_{\hat{\rho}} \; \exp\left\{\frac{iN}{24 \pi} \int
  d^2 x \sqrt{-\hat{g}} \, \left[(\hat{\nabla}\phi)^2 + \phi \hat{R}\right] \right\}\;,
\end{equation}
where the new metric $\hat{g}_{\mu\nu}$ with scale factor
$\hat{\rho}\equiv \rho - \phi$ is used everywhere in the right-hand
side. We obtain,
\begin{equation}
  \label{eq:13}
  Z_{RST} = \int [d\hat{\rho} \, d\hat{\phi} \, db \, dc \,
    d\hat{f}]_{\hat{\rho}} \;\, \mathrm{e}^{iS_{JT}[\hat{\rho},\, \hat{\phi},\, \hat{f}]
    + iS_{gh}[b,\, c]} = Z_{JT}\;,
\end{equation}
where 
\begin{equation}
  \label{eq:14}
  S_{JT} =\int d^2x\,\sqrt{-\hat{g}}\left(\hat{\phi} \hat{R} +
    4\lambda^2 -\frac12\sum_{j=1}^N(\hat{\nabla} \hat{f}_j)^2\right)
\end{equation}
is the action of the flat-space JT model with metric $\hat{g}_{\mu\nu}$, dilaton
$\hat{\phi}$ and matter fields $\hat{f}_j$, see~\cite{Jackiw:1984je,
  Teitelboim:1983ux, Cangemi:1992bj, Afshar:2019axx}.

The equivalence (\ref{eq:13}) is very natural. One can
check~\cite{Cangemi:1992bj} that the classical part of the RST action
is related to the classical flat-space JT gravity by the
transformation\footnote{In the classical case one omits the
  second term in the expression for 
  $\hat{\phi}(\phi)$.}~(\ref{eq:5}). Thus, any quantization on the RST
side can be performed in JT terms and vice  versa. Moreover, the RST
counter-term in Eq.~(\ref{eq:6}) was originally
introduced~\cite{Russo:1992ax} to extend the symmetry  ${\hat{\phi}
  \to \hat{\phi} +   \mbox{const}}$ to one-loop level. This shift
symmetry is manifest in  (\ref{eq:14}), but nonlinearly realized in the
RST terms. It is not a wonder that a consistent quantization
preserving the symmetry reproduces the quantum flat-space JT model.

We summarize that the quantum RST model defined in~(\ref{eq:12}) is a
disguised version of flat-space JT gravity. Note that the JT metric
$\hat{g}_{\mu\nu}$ is flat on the classical field equations 
because variation of (\ref{eq:14}) with respect to $\hat{\phi}$ gives
$\hat{R}=0$. This property guarantees that the Hawking flux of
unphysical fields is zero.

In what follows we will strongly rely on the
fact~\cite{Dubovsky:2017cnj, Dubovsky:2018bmo} that the flat-space JT
gravity is a healthy quantum theory with unitary ${\cal S}$-matrix. This 
implies, in particular, that the RST model remains local and
diffeomorphism-invariant after distortion of the functional measures
performed in~(\ref{eq:12}). Indeed, inverse Weyl
transformation to Eq.~\eqref{eq:11},    
\begin{equation}
  \label{eq:1}
  [d\rho \, d\phi \, db \, dc]_{\rho-\phi} = [d\rho\, d\phi \,
    db \, dc]_{\rho} \, \exp\left\{ \frac{i}{\pi} \int d^2 x
  \sqrt{-g} \left[R \phi - (\nabla \phi)^2\right]\right\}\;,
\end{equation}
gives path integral with canonical functional measures and new
local  counter-terms in the action.

%%%%%%%%%%%%%%%%%%%%%%%%%%%%%%%%%%%%%%
\subsection{Adding the boundary}
\label{sec:one-loop-corrections}
Let us explicitly show that the transformation between the JT and
RST models cannot be performed globally. Consider the classical 
JT vacuum in flat light-cone coordinates~$(u,\, v) =  (t-x,\, t+x)$,
\begin{equation}
  \label{eq:2}
  \hat{g}_{\mu\nu} = \eta_{\mu\nu}\;, \qquad \qquad \hat{\phi} =
  -\lambda^2 uv\;.
\end{equation}
The dilaton $\hat{\phi}$ takes arbitrary
values in this two-dimensional spacetime. On the other hand, the
function $\hat{\phi}(\phi)$ in Eq.~(\ref{eq:5}) is bounded from
below by the critical value $\hat{\phi}_{cr}$ given in the
Introduction: $\hat{\phi}(\phi) \geq \hat{\phi}_{cr}$. Thus, the
transformation~(\ref{eq:5}) cannot be performed at $\hat{\phi}<
\hat{\phi}_{cr}$ i.e.\ in the region between the dashed lines in
Fig.~\ref{fig:intro}b. The latter lines are the true  singularities of
the RST fields.

To remedy the RST model, we introduce a reflective spacetime
boundary along the line of constant dilaton field $\phi
= \phi_0$, $\phi_0<\phi_{cr}$. To this end we restrict all 
fields in the path integral~(\ref{eq:12}) to the submanifold $\phi 
\leq \phi_0$ (the rightmost region in Fig.~\ref{fig:intro}b) and
add the boundary term to the action~\cite{Fitkevich:2017izc, Eremeev},
\begin{equation}
  \label{eq:3}
  S_{RST,\, b} = \int\limits_{\phi=\phi_0}d\tau\left[2K \left(e^{-2\phi_0}
-Q^2\phi_0\right)-\mu\right]\;,
\end{equation}
where $\tau$ is the proper time at $\phi = \phi_0$ and $K = \nabla_\mu
n^\mu$ is the extrinsic curvature computed with the outer
normal $n^\mu$. Expression~(\ref{eq:3}) includes the Gibbons-Hawking
term~\cite{Gibbons:1976ue, Poisson} for the curvature part of the
action (\ref{eq:6}) and a negative ``mass'' $\mu = - 2
\lambda (2 e^{-2\phi_0} +Q^2)$. We will see that the latter parameter
stabilizes the position of the boundary in vacuum,
cf.~\cite{Fitkevich:2017izc}. As before, $Q^2\equiv N/(48\pi)$ appears
in front of all counter-terms.

Note that the boundary makes the RST model weakly coupled, and
the parameter $\mathrm{e}^{2\phi_0} \ll 1$ controls
semiclassical expansion. Indeed, the change of variables
${\phi' \equiv \phi - \phi_0}$ and $f_j' \equiv \mathrm{e}^{\phi_0}\,
f_j$ brings $\mathrm{e}^{-2\phi_0}$ in front of the
action~(\ref{eq:6}), (\ref{eq:3}) and makes all counter-terms
proportional to $Q^2\mathrm{e}^{2\phi_0} \sim
N\mathrm{e}^{2\phi_0}$. Below we are interested in the regime
\begin{equation}
  \label{eq:4}
  \mathrm{e}^{-2\phi_0} \gg Q^2 \gg 1
\end{equation}
where the quantum corrections are small and dominated by loops of matter
fields.

To describe evaporating black holes in the RST model, we
need to derive one-loop Polyakov action with correct boundary
terms. It is uniquely fixed by the path integral
  (\ref{eq:12}). Note first that the boundary introduces 
local and covariant deformation of the JT action: 
performing the transformation (\ref{eq:5}) in Eqs.~(\ref{eq:6}),
(\ref{eq:3}) and adding the correction\footnote{One introduces the
  Gibbons-Hawking term $4i Q^2\int   d\hat{\tau} \, \phi \hat{K}$ in
  the exponent of~(\ref{eq:11}) for consistency.}~(\ref{eq:11}), one obtains
Eq.~(\ref{eq:14}) with the boundary term 
\begin{equation}
  \label{eq:9}
  S_{JT,\, b} = \int\limits_{\hat{\phi} = \hat{\phi}_0} d\hat{\tau}
  \left[ 2 \hat{\phi}_0 \hat{K} - \mu \mathrm{e}^{\phi_0}\right]\;,
\end{equation}
where $\hat{\phi}_0$ is related to $\phi_0$ by Eq.~(\ref{eq:5}). Next,
we recall that Weyl transformation of the Polyakov action is fixed
by the Wess-Zumino 
condition~\cite{Polchinski:1998rq}. Namely, ${\delta_W
\hat{g}_{\mu\nu} = 2 w(x) \, \hat{g}_{\mu\nu}}$ should lead to
\begin{equation}
  \label{eq:15}
  \delta_W S_{JT,\, \mathrm{1-loop}} = \frac{c}{24\pi}\Bigg[  \int
  d^2x \sqrt{-\hat{g}}\,  \hat{R} \, w + 2\int\limits_{\hat{\phi} =
    \hat{\phi}_0} d\hat{\tau}\;  \hat{K} \,w \Bigg]\;,
\end{equation}
in any consistent quantum theory, e.g.\ in the flat-space JT gravity. From now on, we
use $c=N$ because non-dynamical fields will not contribute into the
final Polyakov action anyway.

Solution of Eq.~(\ref{eq:15}) can be conveniently written in terms of
an auxiliary field $\hat{\chi}(x)$ satisfying
\begin{equation}
  \label{eq:16}
  \hat{\Box}\hat{\chi} = -Q\hat{R}\;, \qquad \qquad \hat{n}^\mu
  \hat{\nabla}_\mu \hat{\chi} = 2Q\hat{K}\;.
\end{equation}
The Polyakov action is then
\begin{equation}
  \label{eq:17}
  S_{JT,\, \mathrm{1-loop}} = \int d^2 x \sqrt{-\hat{g}}
  \left[-\frac12 (\hat{\nabla} \hat{\chi})^2 + Q\hat{R}\hat{\chi} \right] +
  2Q\int\limits_{\hat{\phi} = \hat{\phi}_0} d\hat{\tau}\,  \hat{K}\, \hat{\chi}\;.
\end{equation}
One can explicitly check that it satisfies Eq.~(\ref{eq:15}) and
coincides with Eq.~(\ref{eq:10}) in the bulk. Equations~(\ref{eq:16})
simply mean that the action is extremal with respect
to~$\hat{\chi}$. Since we are going to solve the semiclassical field 
equations anyway, we will treat~$\hat{\chi}$ on equal grounds with
other fields.

Rewriting the effective action~(\ref{eq:14}), (\ref{eq:9}),
  (\ref{eq:17}) in the RST terms~(\ref{eq:5}), one obtains 
a consistent Polyakov correction to the RST
model\footnote{Of course, Eq.~(\ref{eq:18}) satisfies Wess-Zumino
  condition by itself, and one can derive it without resorting to the
  flat-space JT model. We imposed consistency requirement in explicitly sane
  terms to avoid confusion.}  
\begin{equation}
  \label{eq:18}
  S_{RST,\, \mathrm{1-loop}} = \int d^2 x \sqrt{-g}
  \left[-\frac12 (\nabla \chi)^2 + QR\chi \right] +
  2Q\int\limits_{\phi = \phi_0} d\tau\,  K\, \chi\;,
\end{equation}
where the auxiliary field $\chi \equiv \hat{\chi} + 2Q \phi$
extremizes the effective action.

To summarize, we have defined the quantum RST model,
  Eqs.~(\ref{eq:12}), (\ref{eq:6}), (\ref{eq:3}), in the most
sensible way. First, we related it locally to the quantum flat-space JT
theory~\cite{Jackiw:1984je, Teitelboim:1983ux} with unitary ${\cal
    S}$-matrix~\cite{Dubovsky:2017cnj, Dubovsky:2018bmo}. Second, we
regularized the singularity of the Weyl transformation with
a dynamical boundary. This gave us the unique one-loop
action~(\ref{eq:18}). In what follows we describe black hole
evaporation using this action.

%%%%%%%%%%%%%%%%%%%%%%%%%%%%%%%%
\section{Evaporating black holes}
\label{sec:rst-model}

%%%%%%%%%%%%%%%%%%%%%%%%%%%%
\subsection{Semiclassical equations}
\label{sec:cauchy-problem}

\begin{sloppy}

Given the equivalence between the quantum RST and JT models, it is
worth reanalyzing semiclassical evaporation of the RST black holes,
cf.~\cite{Russo:1992ax, Russo:1992yh, Chung:1993rf, Strominger:1994xi,
  Das:1994yc}. To this end we consider the one-loop
effective action
\begin{equation}\label{eq:full-action}
S=S_{RST}+S_{RST,\, b}+S_{RST,\, 1-\mathrm{loop}}\;,
\end{equation}
where the bulk action $S_{RST}$, boundary term $S_{RST,\, b}$, and
one-loop correction $S_{RST,\, 1-\mathrm{loop}}$ are given by
Eqs.~(\ref{eq:6}), (\ref{eq:3}), and (\ref{eq:18}), respectively. In
what follows we use only one classical field $f=f_1$ keeping $f_j=0$
at $j\geq 2$. Recall that nevertheless, all fields fluctuate and
contribute into $\chi$. 

\end{sloppy}

Let us review the semiclassical solutions extremizing the effective action; we
leave details of their derivation to Appendices~\ref{sec:derivation}
and~\ref{sec:boundary-eq}. As usual, equations for the propagating
fields $f$, $\chi$ simplify in 
the  conformal frame~\eqref{eq:7} with $ds^2 = -\mathrm{e}^{2\rho} du
dv$, where we use the light-cone coordinates $u$ and $v$.  General solution to these
equations has the form,
\begin{align}
f=f_{\mathrm{in}}(v)+f_{\mathrm{out}}(u)\;, \qquad 
\chi=2Q\rho(u,\, v)+\chi_{\mathrm{in}}(v)+\chi_{\mathrm{out}}(u)\;, 
\label{eq:f_chi_solution}
\end{align}
where $f_{\mathrm{in, out}}$ are the incoming and outgoing wave
packets, while $\chi_{\mathrm{in, out}}$ are their quantum counterparts.

It is worth noting that Eq.~(\ref{eq:7}) does not completely fix the
reparametrization invariance leaving residual conformal symmetry
$u\to\tilde{u}(u)$, $v\to
\tilde{v}(v)$~\cite{Strominger:1994xi}. Quantum wave packets transform
nontrivially under this symmetry, 
\begin{equation}
  \label{eq:31}
  \chi_{\mathrm{in}} \to \tilde{\chi}_{\mathrm{in}}(\tilde{v}) =
  \chi_{\mathrm{in}}(v) + Q \log \left(\partial_v\tilde{v} \right)\;,
  \;\;   \chi_{\mathrm{out}} \to
  \tilde{\chi}_{\mathrm{out}}(\tilde{u}) = \chi_{\mathrm{out}}(u) + Q
  \log \left(\partial_u\tilde{u} \right)\;, 
\end{equation}
see Eq.~(\ref{eq:f_chi_solution}). To explain the transformation law,
we introduce the energy fluxes: the incoming flux
\begin{equation}\label{eq:27}
T_{vv}(v)=(\partial_v f_{\mathrm{in}})^2+
(\partial_v\chi_{\mathrm{in}})^2+2Q\,\partial_v^2\chi_{\mathrm{in}}
\end{equation}
and the outgoing flux $T_{uu}(u)$~--- by the same expression
with $v\to u$ and ``in''~$\to$~``out,'' see Appendix
\ref{sec:derivation} for details. According to Eq.~(\ref{eq:31}), the
fluxes have the standard conformal transformation
laws\footnote{Note that $T_{vv}$ and $T_{uu}$ are not the
  components of a reparametrization-covariant tensor, see their
  definition in Appendix~\ref{sec:derivation}.}, e.g.
\begin{equation}\label{eq:v-conformal-transformation} 
\tilde{T}_{\tilde{v}\tilde{v}}(\tilde{v})=
(\partial_v\tilde{v})^{-2} \left(
T_{vv}(v)+2Q^2\{\tilde{v};v\}\right)\;,
\end{equation}
where the term with the Schwarzian
$\{\tilde{v};v\}=\partial_v^3\tilde{v}/\partial_v\tilde{v}
-\frac{3}{2}(\partial_v^2\tilde{v})^2/(\partial_v\tilde{v})^2$ comes from the
quantum field $\chi_{\mathrm{in}}$. As expected, this term
is proportional to the total central charge $Q^2 \propto N$. 

Now, recall the correspondence (\ref{eq:5}) with the flat-space JT model. Since
the JT metric is flat, $\hat{R}=0$, one can introduce flat coordinates
with $\hat{g}_{\mu\nu} = \eta_{\mu\nu}$. In RST terms this 
corresponds to choosing the ``Kruskal'' gauge with
\begin{equation}
  \label{eq:34}
  \rho = \phi
\end{equation}
in Eq.~\eqref{eq:7}. In the model with a boundary the ``Kruskal''
coordinates $u$ and $v$ have semi-infinite ranges $-\infty <
u < 0$ and $0 < v< +\infty$, see Fig.~\ref{fig:intro}b and 
Appendix~\ref{sec:derivation}. Importantly, equations for the
gravitational sector simplify in these coordinates giving
\begin{equation}\label{eq:29}
\hat{\phi} = e^{-2\phi}+Q^2\phi=-\lambda^2uv +g(v)+h(u)\;, 
\end{equation}
where 
\begin{align}\label{eq:30}
  & g(v)= \frac12 \int\limits_0^v dv' \int\limits_{v'}^{\infty} dv''\left( T_{vv}(v'') +
\frac{Q^2}{(v'')^2}\right) - \frac{Q^2}{2} \,\log (\lambda v)\;, \\
\label{eq:35s}
&h(u)= -\frac12 \int\limits_{-\infty}^u du' \int\limits_{-\infty}^{u'} du''\left( T_{uu}(u'') +
\frac{Q^2}{(u'')^2}\right) - \frac{Q^2}{2} \,\log (-\lambda u) 
\end{align}
are the second primitives of the energy fluxes. 

Let us define the RST vacuum as a solution with ${f_{\mathrm{in}} =
  f_{\mathrm{out}} = 0}$,
\begin{equation}
  \label{eq:36}
  \chi_{\mathrm{in}} = Q \log (\lambda v)\;, \qquad
  \chi_{\mathrm{out}} = Q \log (-\lambda u)\;, \qquad
    \phi = -\frac12 \log \left( -\lambda^2 uv\right)\;.
\end{equation}
This configuration does not look like a vacuum in the ``Kruskal''
coordinates. However, transforming it
via~Eqs.~\eqref{eq:31}, (\ref{eq:v-conformal-transformation}) to the uniformly
accelerating frame with respect to $u$ and~$v$,
\begin{equation}
  \label{eq:37}
  \bar{u} = -\frac{1}{\lambda} \, \log (-\lambda u)\;, \qquad \qquad
  \bar{v} = \frac{1}{\lambda} \, \log (\lambda v)\;.
\end{equation}
one finds explicitly flat empty spacetime with $ds^2 =
-d\bar{u}d\bar{v}$, $\phi = \lambda (\bar{u}-\bar{v})/2$, ${f =
  \chi = 0}$, and $\bar{T}_{\bar{u}\bar{u}} = \bar{T}_{\bar{v}\bar{v}}
= 0$. Thus, from the viewpoint of the quantum JT model $u$ and
$v$ are flat coordinates and the RST vacuum (\ref{eq:36}) is a Rindler
state. For the semiclassical RST model, on the contrary, the RST
vacuum~(\ref{eq:36}) is the only empty flat solution and $\bar{u} =
\bar{t} - \bar{x}$, $\bar{v} = \bar{t}+\bar{x}$ are flat
coordinates. Note that the vacuum fluxes are negative in the
``Kruskal'' gauge: $T_{vv} = -Q^2/v^2$ and $T_{uu} = -Q^2/u^2$. They
are subtracted in the integrands of Eqs.~\eqref{eq:30} and
\eqref{eq:35s} for convergence.

In what follows we describe the RST boundary $\phi = \phi_0$ by a
function ${u = U(v)}$. We derive the reflection laws on this boundary in
Appendix~\ref{sec:boundary-eq} by  extremizing the effective
action~\eqref{eq:full-action} with respect to the boundary values of
the matter fields. This gives,
\begin{equation}\label{eq:20}
f_{\mathrm{out}}(U(v)) = f_{\mathrm{in}}(v)\;, \qquad\qquad 
\chi_{\mathrm{out}}(U(v))=\chi_{\mathrm{in}}(v)+Q\log(\partial_v U(v))\;,
\end{equation}
implying that the boundary acts precisely like a conformal
transformation from $v$ to $u = U(v)$, cf.\ Eq.~(\ref{eq:31}). The
reflection laws relate the energy fluxes,
\begin{equation}\label{eq:23}
T_{uu}(U(v))=(\partial_v U)^{-2}\left(T_{vv}(v)+2Q^2\{U;\,v\}\right)\;.
\end{equation}
cf.\ Eq.~(\ref{eq:v-conformal-transformation}). Note that the
boundary condition~(\ref{eq:23}) 
generically appears in all models with moving mirrors,
cf.~\cite{Davies:1976hi, Wilczek:1993jn, Chung:1993rf}, where the
first term conserves classical energy in the rest frame of the
mirror and the Schwarzian represents quantum particle
production.
 
Equation of motion for the boundary $u = U(v)$ is obtained by
recalling that $\phi =\phi_0$ along this line, 
\begin{equation}\label{eq:boundary_ode_wz}
\partial_v U=
e^{2\phi_0} \, \frac{q^2}{\lambda^2}
\left(\partial_vg-\frac{Q^2}{2}\frac{\partial_v^2U}{\partial_vU}
-\lambda^2U\right)^2\;,
\end{equation}
where $q=(1+e^{2\phi_0}Q^2/2)^{-1}$, see Appendix
\ref{sec:boundary-eq}. We stress that the derivation of 
this equation heavily relies on the reflection law (\ref{eq:23});
modification of the latter changes $U(v)$ as well. Terms with $Q^2$ in
Eq.~(\ref{eq:boundary_ode_wz}) represent quantum corrections. They are
small in the semiclassical regime $Q^2 \ll \mathrm{e}^{-2\phi_0}$
because after rescaling $f = \mathrm{e}^{-\phi_0} f'$,
$\chi = Q \chi'$, and $U = \mathrm{e}^{-2\phi_0}
U'$ equation~(\ref{eq:boundary_ode_wz}) involve $Q$ and
$\phi_0$ in  the combination $Q^2 \mathrm{e}^{2\phi_0} \ll 1$. 

Solving Eq.~(\ref{eq:boundary_ode_wz}) for the RST
vacuum~\eqref{eq:36}, one obtains, 
\begin{equation}\label{eq:static-boundary} 
U(v)=-e^{-2\phi_0}/(\lambda^2v)\;.
\end{equation}
Thus, the vacuum boundary is static, $\bar{x}(\bar{t}) =
-\phi_0/\lambda$, cf.\ Eq.~\eqref{eq:37}.  

In what follows we solve the Cauchy problem in the semiclassical RST
model. Namely, we prepare the finite-energy incoming 
wave packets $f_{\mathrm{in}}(v)$ keeping the initial quantum field
in vacuum, ${\chi_{\mathrm{\mathrm{in}}} = Q \log (\lambda
  v)}$. Computing the incoming flux~\eqref{eq:27}, \eqref{eq:30}, we
solve the equation of motion (\ref{eq:boundary_ode_wz}) for 
$U(v)$. Then the reflection law (\ref{eq:23}) determines the outgoing
flux $T_{uu}$, while Eqs.~(\ref{eq:35s}), \eqref{eq:29} fix the
spacetime geometry~${\rho = \phi}$.

%%%%%%%%%%%%%%%%%%%%%%%%%%%%%%%%%%%
\subsection{General properties of solutions}
\label{sec:semicl-solutions}
\begin{figure}
  
  \hspace{2cm}(a) \hspace{5cm}(b) \hspace{5cm}(c)

  \vspace{-5mm}
  \centerline{\includegraphics{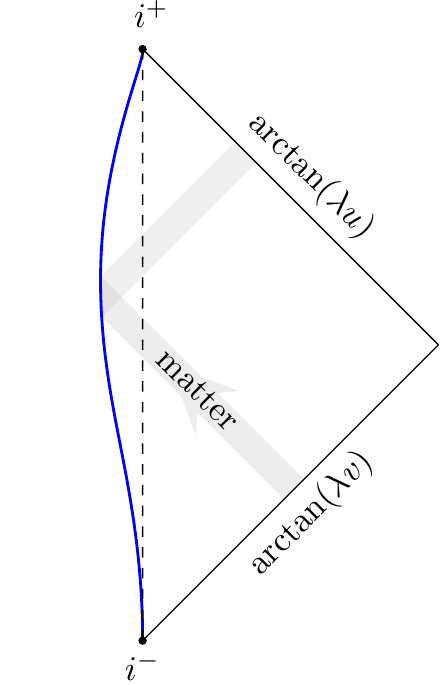}
    \includegraphics{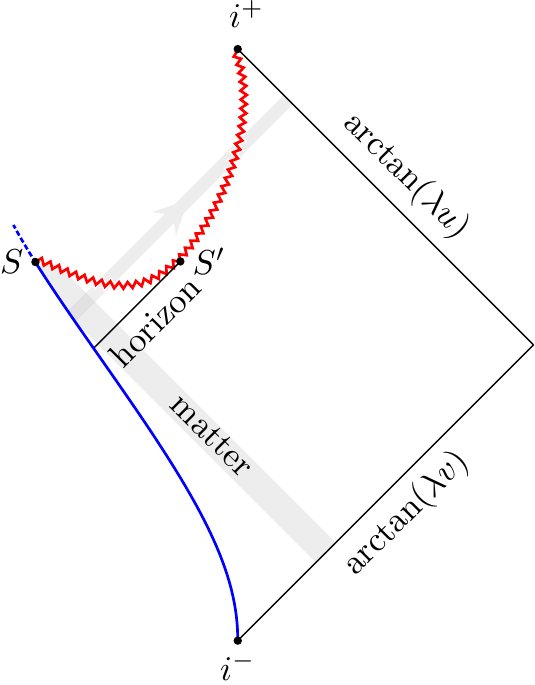}
    \includegraphics{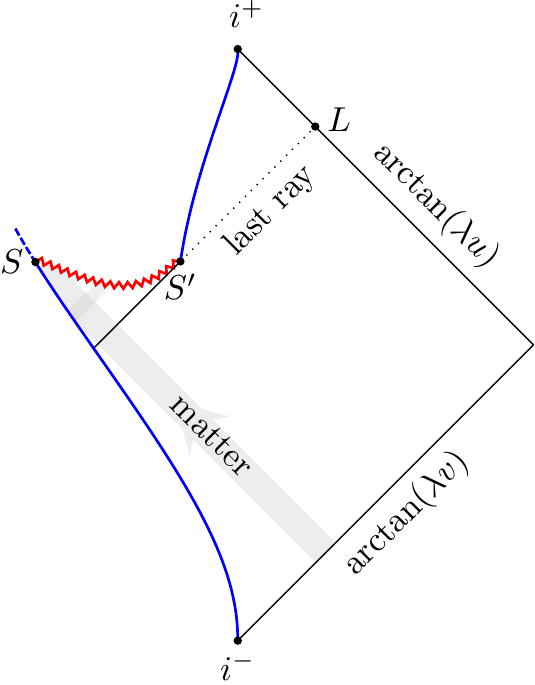}
  }
  \caption{Numerical evolution of the wave packet~(\ref{eq:40})  in
    the one-loop RST model; the respective parameters are $Q  = 0.5$,
    $\sigma = 8.5 \lambda$, and $v_0 = 0.7 /\lambda$. For
    visualization purposes we set $\phi_0=0$. (a) Low-energy
    reflection at $A = 2.5\, \lambda^2$. (b) Black hole formation
      and evaporation
    at high energies: $A = 10\, \lambda^2$. (c)~Natural extension of
    the solution~(b). \label{fig:BH-exact-numerical}
  }
\end{figure}

To warm up, we explore the semiclassical solutions numerically. We
fix the incoming wave packet, 
\begin{equation}
  \label{eq:40}
  (\partial_v f_{\mathrm{in}})^2 = A \, \cosh^{-2}\left(\sigma( v -
  v_0)\right)\;,
\end{equation}
keeping its quantum counterpart $\chi_{\mathrm{in}}$ in
vacuum~\eqref{eq:36}. Here the parameters $A$, $\sigma$, and $v_0$
represent respectively the amplitude, width, and initial
position of the wave packet. Numerical integration of
Eq.~(\ref{eq:boundary_ode_wz}) gives the boundary trajectories
$U(v)$ at low and high energies of $f_{\mathrm{in}}$, see the leftmost
solid lines in Figs.~\ref{fig:BH-exact-numerical}a
and~\ref{fig:BH-exact-numerical}b, respectively.

The low-energy solution in Fig.~\ref{fig:BH-exact-numerical}a
describes trivial reflection of matter. Indeed, the respective
boundary starts at $i^-$ close to the vacuum solution $U\propto
v^{-1}$ (dashed line), departs from it due to matter pressure
and approaches it again at $i^+$. As a direct consequence,
the reflected energy flux $T_{uu}$ in Eq.~\eqref{eq:23} tends to the
vacuum value at $u\to \pm \infty$, and the the spacetime~\eqref{eq:29}
becomes flat in the asymptotic past and future. We argue in
Appendix~\ref{sec:energy-conservation} that the total energy of the
outgoing matter in this case coincides with the energy of
$f_{\mathrm{in}}$.

The solution changes completely if the incoming energy exceeds certain
value, see Fig.~\ref{fig:BH-exact-numerical}b.  In this case the
boundary crosses the axis $u=0$ and continues growing. Numerically
computing the dilaton in Eq.~\eqref{eq:29}, we discover another
anomaly. Namely, the line $\phi(u,\, v) = \phi_0$ has two
intersecting branches: the original timelike boundary $U(v)$ and
the additional line $S i^+$ with spacelike part $SS'$. The
semiclassical boundary conditions~\eqref{eq:20} are not satisfied
along $Si^+$, and one cannot enforce them without
strongly violating causality. As a consequence,
this line plays the role of a singularity in the
  one-loop RST model. Recall that we restrict the spacetime to the
region $\phi < \phi_0$, below the line $Si^+$. Yet, the matter
freely crosses this line and goes away, see
Eq.~\eqref{eq:f_chi_solution}. At late times the singularity
becomes timelike (part $S' i^+$ in
Fig.~\ref{fig:BH-exact-numerical}b). We will consider this region
in Sec.~\ref{sec:beyond-endpoint-1}.

Let us show that the salient features of solutions in
Figs.~\ref{fig:intro}c and \ref{fig:BH-exact-numerical}a,b are, in fact,
generic. First, the RST vacuum is indeed stable, and all low-energy
solutions correspond to reflection. Adding small perturbation
$\delta \bar{u} \ll \lambda^{-1}$ to the vacuum solution,
$U=-e^{-2\phi_0}(1 - \lambda \delta \bar{u})/(\lambda^2v)$, one arrives
to the linearized equation,
\begin{equation}\label{eq:linear-ode} 
Q^2 \partial_{\bar{v}}^2 \delta \bar{u} + \left(\mathrm{e}^{-2\phi_0}
- Q^2 / 2\right) (\lambda \partial_{\bar{v}} + \lambda^2) \delta \bar{u} =
2\partial_{\bar{v}} g + \lambda Q^2\;,
\end{equation}
where the asymptotic RST coordinate $\bar{v}$ is used,
Eq.~\eqref{eq:37}. Both fundamental solutions $\delta \bar{u} 
\propto \exp(ik \bar{v})$ of the left-hand side in this equation
have $\mathrm{Im}\, k > 0$ and therefore
die off exponentially in the asymptotic future if $Q^2
\mathrm{e}^{2\phi_0} < 2$, cf.~\eqref{eq:4}. This means that the
boundary shifted by the matter source in the right-hand side always
returns back to the vacuum position, describing reflection.  This regime holds if
the incident energy is below some threshold. 

\begin{sloppy}

Second, we prove that the spacelike ``singularity'' $\phi =
  \phi_0$ always forms in the limit of high matter flux, cf.\
  Fig.~\ref{fig:BH-exact-numerical}b. In this case the function
$u=U(v)$ grows fast and crosses the axis $u=0$ due to large right-hand
side in Eq.~\eqref{eq:boundary_ode_wz}. Define the apparent
horizon\footnote{Here we treat $\exp(-2\phi)$ as an
    analog of the sphere area in the multidimensional gravity.} $u_{a}(v)$ as
the boundary of the region where the curves of constant $\phi$
become space-like. Taking the derivative of Eq.~\eqref{eq:29}
along these curves and finding the point $du/dv=0$, one obtains
$u_a=\partial_vg(v)/\lambda^2$. Notably, the apparent horizon
approaches the axis $u=0$ at $v\to +\infty$, see Eq.~\eqref{eq:30}. Then
the growing boundary $u = U(v)$ intersects $u_a(v)$ at some point $S$:
$U(v_s) = u_a (v_s)$, and this is  where the
singularity appears. Indeed, $\partial_u\phi$ and $\partial_v \phi$
are zero at $S$, see Eqs.~\eqref{eq:29} and~\eqref{eq:32}. Hence, 
Taylor series expansion for $\phi(u,\, v) - \phi_0$ starts from
quadratic terms near $S$. It will be convenient to relate the unknown
coefficients of this expansion to the apparent velocity of the
boundary $u_1 = \partial_v U(v_s)$ using $\phi(U(v),\, v) =
\phi_0$. Equation $\phi = \phi_0$ in the vicinity of the point $S$
takes the form,
\begin{equation}
  \label{eq:49}
  u_1^2 T_{vv} (v - v_s)^2  - (T_{vv} + 4\lambda^2 u_1)  (u - u_s)^2
  + 4\lambda^2 u_1^2 (u - u_s) (v - v_s) = 0\;,
\end{equation}
where $T_{vv}$ is the incoming flux at $v = v_s$ and we omitted
higher-order terms in $v-v_s$ or $u -
  u_s$. Equation~(\ref{eq:49}) has two solutions: the time-like
boundary $u - u_s = u_1 (v - v_s)$, where $u_1 > 0$ due
to Eq.~\eqref{eq:boundary_ode_wz}, and  the spacelike branch ${u - u_s = -u_1
T_{vv} (v - v_s)/(T_{vv} + 4\lambda^2 u_1)}$. This proves that the
space-like singularity $\phi = \phi_0$ generically appears in
high-energy solutions.

\end{sloppy}

Note that the region behind the light-like ``horizon'' in
Fig.~\ref{fig:BH-exact-numerical} can be interpreted as the black hole
interior, since matter in this region cannot escape the
singularity.

%%%%%%%%%%%%%%%%%%%%%%%%%%%%%%%%%%%%%%%%

\subsection{Solvable deformation}\label{sec:solv-def}

\begin{figure}
  \centerline{\includegraphics{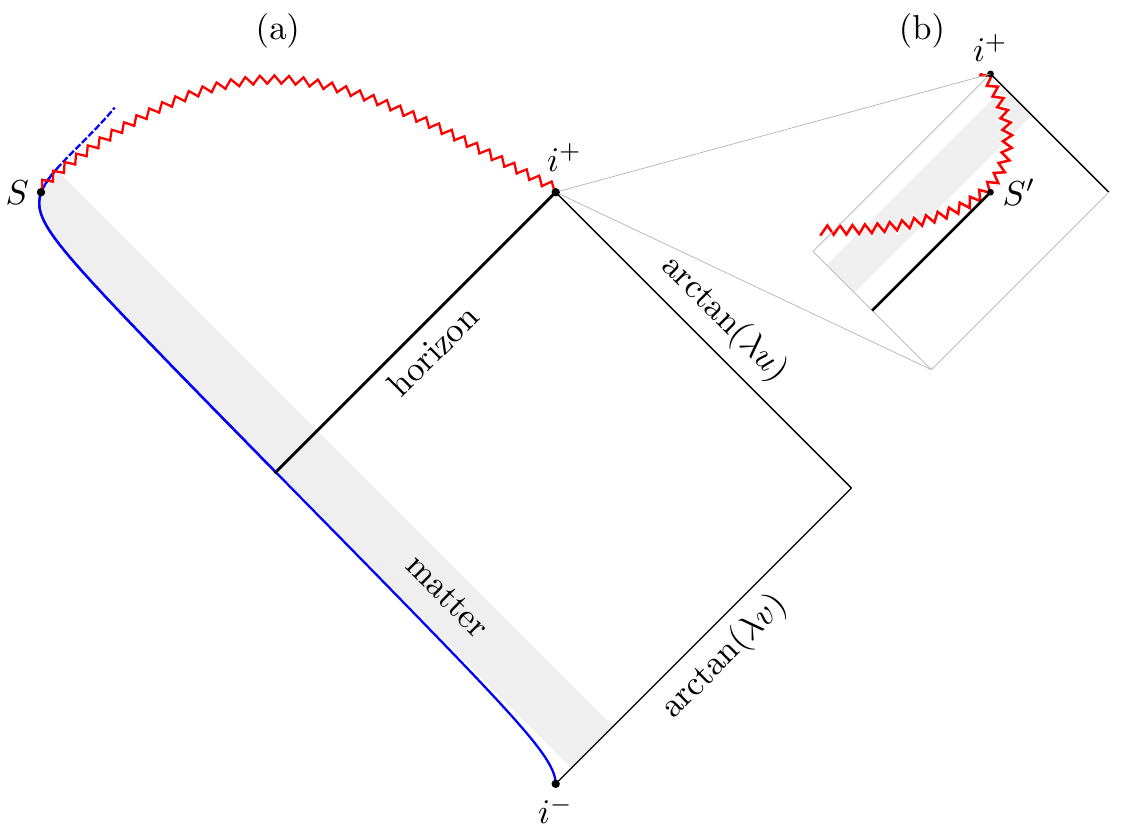}}
  \caption{
    (a)~Analytic solution~\eqref{eq:54}, \eqref{eq:55}, \eqref{eq:56}
    describing black hole formation and evaporation in the modified
    semiclassical model, cf.\ Fig.~\ref{fig:BH-exact-numerical}b.
    For demonstration purposes we use $a=0$, $b=\lambda^{-1}$,
    $\phi_0=0$, and $Q=1$. (b)~Zoom-in on the final stage of
    evaporation in Fig.~(a). 
    \label{fig:BH_numerical}
  }
\end{figure}

Unfortunately, Eq.~(\ref{eq:boundary_ode_wz}) is not exactly
solvable. The method to avoid this obstacle was suggested
  in~\cite{Eremeev}. To this end one imposes the Dirichlet 
boundary conditions\footnote{More precisely, $\chi$ should be constant
  on every simply connected part of the boundary,
  cf.\ Sec.~\ref{sec:beyond-endpoint-1}.} on the auxiliary field
$\chi$, 
\begin{equation}
  \label{eq:51}
  \chi = \mbox{const} \qquad \mbox{along} \qquad u = U(v)\;.
\end{equation}
This modification is healthy at the one-loop level, but not in
the full quantum approach: recall that the boundary
condition for $\chi$ is fixed by the Wess-Zumino consistency
condition~\cite{Polchinski:1998rq}. Nevertheless, we will see
that solutions in the deformed model~(\ref{eq:51}) approximate well
the original ones.

\begin{sloppy}

  The condition (\ref{eq:51}) gives the simplified reflection law
  for the quantum field, ${\chi_{\mathrm{out}}(U(v)) = - \chi_{\mathrm{in}}(v)+
    \mbox{const}}$, cf.\ Eq.~\eqref{eq:20}. As a consequence,
  the naive energy-momentum tensors $T^{(s)}_{vv} = (\partial_v
  f_{\mathrm{in}})^2 + (\partial_v \chi_{\mathrm{in}})^2$ and
  $T^{(s)}_{uu}$ reflect from the boundary classically,
  cf.~\cite{Russo:1992ax},
\begin{equation}
  \label{eq:50}
(\partial_vU)^2T_{uu}^{(s)}=T_{vv}^{(s)}\;.
\end{equation}
 Physically, this means that the
modification~(\ref{eq:51}) switches off direct particle production  by
the accelerating boundary but leaves the field $\chi$
carrying the Hawking flux.

\end{sloppy}

In Appendix~\ref{sec:solvable-deformation} we derive equation for
$U(v)$ in the solvable model, see also~\cite{Eremeev},
\begin{equation}\label{eq:boundary_ode_sd}
\partial_v U=
e^{2\phi_0}\,\frac{q^2}{\lambda^2}
\left(\partial_vg+\frac{Q^2}{v}-\lambda^2U\right)^2
\end{equation}
where the incoming vacuum $\chi_{\mathrm{in}} = Q \log(\lambda v)$ for the quantum
field  is assumed. Comparing Eqs.~(\ref{eq:boundary_ode_sd}),
(\ref{eq:boundary_ode_wz}), one finds that the modification replaces
$\partial_v^2 U /\partial_v U$  in the original equation 
with $-2/v$. This is trustworthy  
in the limit of high energies when the boundary crosses the horizon
before colliding with $f_{\mathrm{in}}$. Indeed, the solution before
the collision~is~\cite{Chung:1993rf, Eremeev}
\begin{equation}
  \label{eq:57}
        U\approx - \mathrm{e}^{-2\phi_0}/(\lambda^2
        v) + c_1\;, \qquad \mbox{with} \qquad c_1 = \lambda^{-2}
        \lim_{v\to 0}  \left[ \partial_v g(v) + Q^2/(2v)\right]\;.
\end{equation}
With this behavior,
equations~(\ref{eq:boundary_ode_sd}) and (\ref{eq:boundary_ode_wz})
coincide everywhere outside the horizon. Apart from the
high-energy limit, we expect that the model (\ref{eq:51}) will be
useful at the qualitative level for any parameters. 

Importantly, Eq.~(\ref{eq:boundary_ode_sd}) can be solved
exactly~\cite{Fitkevich:2017izc}. Namely, changing the variables,
\begin{equation}
  \label{eq:54}
  U(v) = \frac{1}{\lambda^2} \left\{ \partial_v g + \frac{Q^2}{v} -
  \frac{\mathrm{e}^{-2\phi_0}}{q^2} \, W(v)\right\}\;,
\end{equation}
one obtains a Riccati equation
\begin{equation}
  \label{eq:55}
  (\partial_v f_{\mathrm{in}})^2 =  - \frac{Q^2}{v^2} - 2 \;
  \frac{\mathrm{e}^{-2\phi_0}}{q^2} \left( \partial_v W + W^2\right)\;,
\end{equation}
for the unknown $W(v)$. Alternatively, one can
fix $W(v)$ and compute $U(v)$, $f_{\mathrm{in}}(v)$
by Eqs.~(\ref{eq:54}), (\ref{eq:55}). The nontrivial problem, however, is
to find the $W$-ansatz that gives localized and
positive-definite $(\partial_v f_{\mathrm{in}})^2$.

It was shown in Ref.~\cite{Fitkevich:2017izc} that rational
functions from a wide class satisfy the necessary requirements. The
simplest ansatz is
\begin{equation}
  \label{eq:56}
  W=\frac{q}{v}+\frac1{v-c}-\frac{v-a}{(v-a)^2+b^2}\;,
\end{equation}
where $a$ and $b$ are free real parameters and the constant
\begin{equation}
  c = a + \frac{1}{2(1-q)} \left[\sqrt{a^2 + 4b^2 q(1-q)}  - a\right]\;,
\end{equation}
is fixed by requiring the regularity of the incoming flux: the
pole at $v=c$ in Eq.~(\ref{eq:55}) should be absent. Once this is
fixed, $(\partial_v f_{\mathrm{in}})^2$ is a smooth and
positive-definite function in a wide range of $a$, $b$. Besides,
one can explicitly check that the asymptotic flux
${\bar{T}_{\bar{v}\bar{v}} = \mathrm{e}^{2\lambda \bar{v}} \,
  (\partial_v f_{\mathrm{in}})^2}$ is localized i.e.\ vanishes
exponentially at $\bar{v} \to \pm \infty$, see Eq.~\eqref{eq:37}.
 
Penrose diagram for the exact solution~\eqref{eq:54}, \eqref{eq:55}, 
\eqref{eq:56} is shown in Fig.~\ref{fig:BH_numerical}. It has the 
same qualitative features as the original diagram in
Fig.~\ref{fig:BH-exact-numerical}b, with main distinctions related
to different choice of the incoming flux, cf.\ Eqs.~(\ref{eq:55}),
(\ref{eq:56}) and (\ref{eq:40}). Note that using the method of  
Ref.~\cite{Fitkevich:2017izc} one can construct an infinite number of
multi-parameter solutions with different shapes of the incoming wave
packets. The latter solutions can be classified
using the Gaudin spin chain as an auxiliary tool. 

%%%%%%%%%%%%%%%%%%%%%%%%%%%%%%%%%%%%%
\subsection{Beyond the endpoint}
\label{sec:beyond-endpoint-1}
At late times the singularity in Fig.~\ref{fig:BH-exact-numerical}b
becomes timelike, cf.\ Fig.~\ref{fig:BH_numerical}b. The endpoint $S'
= (u_e,\, v_e)$ where this happens satisfies
\begin{equation}
  \label{eq:59}
  \hat{\phi}(u_e,\, v_e) = \hat{\phi}_0 \;, \qquad\qquad \partial_v
  \hat{\phi}(u_e,\, v_e) \equiv -\lambda^2 u_e + \partial_v g(v_e) = 0\;,
\end{equation}
where we used Eq.~\eqref{eq:29}.

Let us compute $u_e$ and $v_e$  in the
quasi-stationary limit~\eqref{eq:4} when evaporation takes large
asymptotic time $\bar{t}$ due to small  $Q^2$. In this case $v_{e}
\propto \mathrm{e}^{\lambda \bar{t}}$ is exponentially large and $u_e$
is small\footnote{The value of $\lambda u_e$ is visibly large in
  Fig.~\ref{fig:BH-exact-numerical}b where $Q^2 \sim
  \mathrm{e}^{-2\phi_0}$ for visualization purposes. Typical black
  hole spacetimes at small $Q^2$ have $\lambda u_e \ll 1$ like in
  Fig.~\ref{fig:BH_numerical}.},  
see Eq.~\eqref{eq:37}. Thus, the incoming flux $T_{vv}(v_e)$ at
  the endpoint is already in
vacuum implying
\begin{equation}
  \label{eq:60}
  g(v) \approx \frac{E_{\mathrm{in}}}{2\lambda} - \frac{Q^2}{2} \,
  \log(\lambda v) \qquad\qquad \mbox{at}\qquad v\sim v_e\;,
\end{equation}
where we expressed the integral in Eq.~\eqref{eq:30} in terms of  the total
energy $E_{\mathrm{in}}$, see Eq.~\eqref{eq:41} of
Appendix~\ref{sec:energy-conservation}. The other   
integral $h(u)$ is related to $g(v)$ by reflection from
$U(v)$. Introducing the point $v_{\times}$ where
$U(v_\times)=0$, one obtains, 
$h(u_e) \approx \hat{\phi}_0 - g(v_{\times})$. This expression and
Eqs.~(\ref{eq:59}), (\ref{eq:60}), \eqref{eq:29} give the
solution~\cite{Russo:1992ax},
\begin{equation} 
  \label{eq:62}
  u_e \approx - \frac{Q^2}{2\lambda^2 v_e} \;, \qquad \qquad v_e\approx
  \frac{1}{\lambda} \exp\left\{1  + \frac{E_{\mathrm{in}} - 2\lambda
    g(v_{\times})}{\lambda Q^2} \right\}\;,
\end{equation}
which confirms that $v_e$ is indeed exponentially large at small
$Q^2$. Indeed, at large energies when Eq.~\eqref{eq:57} is
approximately valid, $2\lambda g(v_\times)  \approx M_{cr}$, where
$M_{cr} =  2\lambda\mathrm{e}^{-2\phi_0}$ is the minimal black hole
mass at $Q^2 \to 0$, see~\cite{Callan:1992rs, Fitkevich:2017izc}. We
will see that $M_{cr}$ coincides with the black hole mass at the
endpoint; it is smaller than $E_{\mathrm{in}}$.

We have just shown explicitly that the
endpoint~(\ref{eq:62}) exists in generic quasi-stationary spacetimes of
evaporating black holes. Beyond the endpoint the line $\phi =
\phi_0$ is a naked timelike singularity. A natural way to
cure this pathology is to impose reflective boundary
conditions~\eqref{eq:23} at the timelike branch $S'i^+$ of the line
$\phi = \phi_0$.  In this case one solves
Eq.~\eqref{eq:boundary_ode_wz} for the second boundary $U_2(v)$
starting from $U_2(v_e) = u_e$ with some apparent velocity $\partial_v
U_2(v_e)$, see Fig.~\ref{fig:BH-exact-numerical}c. Since the incoming
flux is already at vacuum, $U_2(v)$ trivially approaches the
equilibrium trajectory $U_2 \to - \mathrm{e}^{-2\phi_0} /( \lambda^2 v
)$ at $v\to +\infty$, see discussion in
Sec.~\ref{sec:semicl-solutions}. Nevertheless, the resulting spacetime
is special in two respects. First, it depends on the arbitrary
parameter $\partial_v U_2(v_e)$. Second, it is non-analytic at the
last ray $S'L$ due to the jump of the boundary condition at
$S'$.

\begin{figure}
  % \hrule

  \vspace{1mm}
  \hspace{4.5cm}(a) \hspace{8cm}(b)

  \vspace{-5mm}
  \centerline{\includegraphics{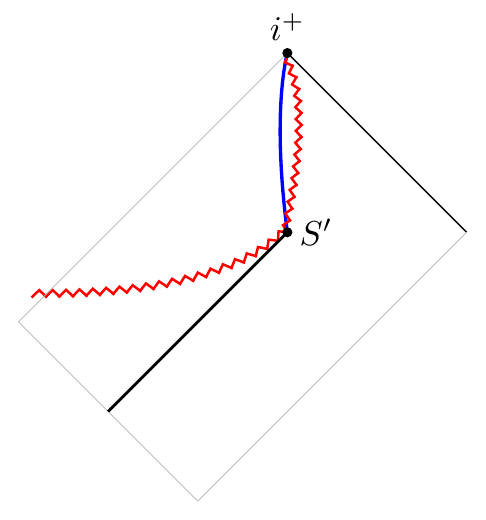}\hspace{2cm}
    \includegraphics{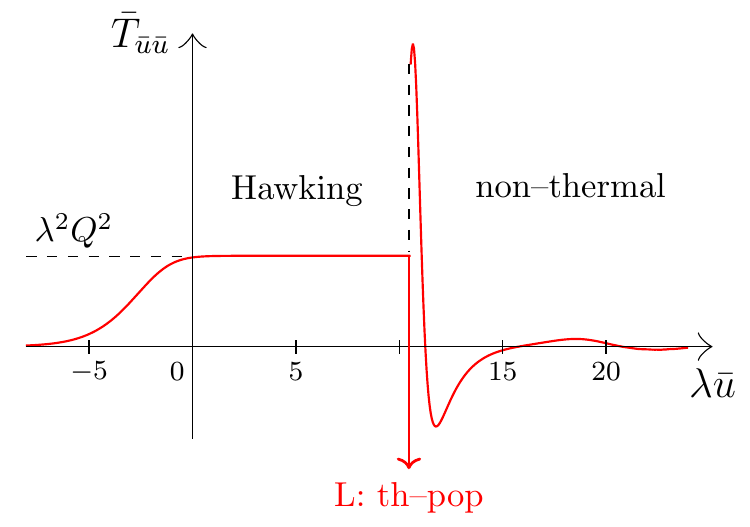}}
  % \hrule
  \caption{(a) Solution with the second boundary violating
    Eq.~\eqref{eq:63},
    cf.\ Fig.~\ref{fig:BH-exact-numerical}c. (b) The asymptotic
    outgoing energy flux for the numerical solution in
    Fig.~\ref{fig:BH-exact-numerical}c.\label{fig:out-flux}}
\end{figure}

Importantly, the initial velocity $\partial_vU_2(v_e)$ of the new
boundary is bounded from above. Indeed, the endpoint is a meeting
place of $U_2(v)$ with the apparent horizon $\partial_v \hat{\phi}
=0$,  see Eq.~(\ref{eq:59}). In Sec.~\ref{sec:semicl-solutions} we learned
that the line $\phi = \phi_0$ generically bifurcates at such points,
with reflective boundary conditions broken along the additional ``singular''
branch. We therefore require that the singularity is not naked like in 
Fig.~\ref{fig:out-flux}a, but hides under $U_2(v)$. This gives
\begin{equation}
  \label{eq:63}
  0 < \partial_v U_2(v_e) < -\frac{T_{vv}(v_e)}{2\lambda^2} \approx
  \frac{Q^2}{4\lambda^2 v_e^2}\;,
\end{equation}
where we solved Eq.~\eqref{eq:49} and substituted
the vacuum flux in the approximate equality. We will see
that Eq.~(\ref{eq:63}) ruins the ``remnant'' explanation of the RST
information paradox, cf.~\cite{Almheiri:2013wka}.

Now, consider the energy balance of the complete solution in
Fig.~\ref{fig:BH-exact-numerical}c. Using
Appendix~\ref{sec:energy-conservation}, we compute the total energy of
black hole emission before the last ray,
\begin{multline}
  \label{eq:64}
  E_{\mathrm{out},\, u < u_e} \approx -2\lambda h(u_e) - \lambda Q^2
  \log (-\lambda u_e) \\= E_{\mathrm{in}} - M_{\mathrm{cr}} - \lambda Q^2
  \log (Q^2 \mathrm{e}^{2\phi_0}) + O(Q^2)\;,
\end{multline}
where we ignored $O(u_e)$ corrections, used Eqs.~\eqref{eq:59}, \eqref{eq:29},
\eqref{eq:60}, \eqref{eq:62} and again introduced the minimal black hole mass
$M_{\mathrm{cr}} = 2\lambda \mathrm{e}^{-2\phi_0}$.

\begin{sloppy}

The outgoing energy~(\ref{eq:64}) appears due to Hawking
effect. Indeed, in the small vicinity of the horizon one can write  $U(v)
\approx \partial_v U(v_\times) \cdot (v - v_\times)$, where
${U(v_\times)=0}$. In terms of the asymptotic coordinate $\bar{u}$ this
vicinity is a large region, 
since ${\bar{u} \approx -\lambda^{-1} \log(\lambda v_\times -
  \mathrm{e}^{\lambda \bar{v}})} + \mbox{const}$ is singular at 
$v\to v_{\times}$. Using this $\bar{u}(\bar{v})$, one computes
the reflected energy flux via Eq.~\eqref{eq:23},
\begin{equation}
  \label{eq:65}
  \bar{T}_{\bar{u}\bar{u}} \approx \frac{\bar{T}_{\bar{v}\bar{v}} \,
  \mathrm{e}^{-2\lambda \bar{u}}}{\lambda^2 v_\times^2}  + \lambda^2 Q^2\;,
\end{equation}
where the second term comes from the
Schwarzian. With time, the stimulated emission in the first term dies
off leaving the stationary thermal flux $\lambda^2 Q^2 =  \lambda^2 N
/(48 \pi)$. Recall that the right-moving Bose gas with $N$ species and
Hawking temperature $T_H=\lambda / (2\pi)$ has 
$$
\mbox{thermal flux}=N\int_0^{\infty}\frac{dk}{2\pi} \,
\frac{k}{e^{k/T_H}-1}=N \, \frac{ \pi T_H^2}{12} = \lambda^2 Q^2\;, $$
precisely the same as in Eq.~\eqref{eq:65}. 

\end{sloppy}

Note that although $h(u)$ is continuous at $u = u_e$, its derivative
is not. Indeed, differentiating~\eqref{eq:29} along the boundaries
$U(v)$ and $U_2(v)$ at $u = u_e \pm 0$, we find,
\begin{equation}
  \label{eq:66}
  u_e\partial_u h (u_e - 0) = O(u_e)\;, \qquad u_e\partial_u h (u_e+0)
  = \lambda^2 u_e v_e = -Q^2/2\;,
\end{equation}
where Eqs.~\eqref{eq:59}, \eqref{eq:62} were used. This jump reflects
$\delta$-function singularity of the outgoing flux along the last ray ---
the thunderpop. Using Eq.~\eqref{eq:41} of
Appendix~\ref{sec:energy-conservation}, we find that the energy of
the thunderpop is negative~\cite{Russo:1992ax},
\begin{equation}
  \label{eq:67}
  E_{\mathrm{th-pop}} = 2\lambda u_e \partial_u h\Big|_{u_e - 0}^{u_e
    + 0} = - \lambda Q^2\;.
\end{equation}
In Sec.~\ref{sec:thund} we will remind that the thunderpop is an
inevitable artifact of non-analytic sewing of two classical spacetimes
along the last ray. It gives negligible energy contribution 
at small $Q^2$, so one can disregard it altogether.

In Appendix~\ref{sec:energy-conservation} we show that the total energy is
conserved for the complete solution in Fig.~\ref{fig:BH_numerical}c
which starts from the RST vacuum in the past and arrives to it in the
future, see also~\cite{Eremeev}. This means that the remaining energy
\begin{equation}
  \label{eq:68}
  E_{\mathrm{out},\, u > u_e} = M_{\mathrm{cr}} + \lambda Q^2 \log (Q^2
  \mathrm{e}^{2\phi_0}) + O(Q^2)
\end{equation}
is emitted\footnote{One can directly calculate it using
  Eq.~\eqref{eq:41} of Appendix~\ref{sec:energy-conservation}.} in
a non-thermal way at $u > u_e$. Expression~\eqref{eq:68} is
the mass of the critical black hole at the endpoint of
  evaporation, decaying afterwards.

In Fig.~\ref{fig:out-flux}b we plot the outgoing energy flux
for the numerical solution in Fig.~\ref{fig:BH-exact-numerical}c. It
displays all features observed above. Notably, it is not strictly
positive-definite beyond the last ray, which is
expected~\cite{Bianchi:2014vea, Bianchi:2014qua, Good:2019tnf}.

%%%%%%%%%%%%%%%%%%%%%%%%%%%%%%%%%
\section{Information loss revisited}
\label{sec:information-loss}

%%%%%%%%%%%%%%%%%%%%%%%%%%%%%%%%%%%%%%% 
\subsection{Endpoint singularity}
\label{sec:ambiguity}

At first glance, the semiclassical RST solutions look smooth~--- even
their ``singularities'' are just the spacelike branches of the line
$\phi = \phi_0$. Nevertheless, one runs into a trouble trying to apply
these solutions to computation of the effective
action~\eqref{eq:full-action}. Indeed, the left border of the
  spacetime in Fig.~\ref{fig:BH-exact-numerical}b is not smooth but
constructed from three time- and spacelike pieces (lines $i^{-}S$,
$SS'$, and $S' i^+$). It is not clear how to evaluate the
extrinsic curvature $K$ and therefore the Gibbons-Hawking
term~\eqref{eq:3} at the angles $S$ and $S'$ of these pieces.

One can try to regularize $S'$ with a smooth curve $U_{reg}(v)$
going between the spacelike and timelike branches of $\phi =
\phi_0$. Then the curve becomes light-like at some point 
$v_0  \approx v_e$, with $U_{reg}(v) - U_{reg}(v_{0}) \propto (v  -
v_{0})^2$ near it. Using Appendix~\ref{sec:boundary-eq},   
we evaluate the extrinsic curvature of $U_{reg}(v)$ in this region, 
\begin{equation}
  \label{eq:44}
  K(\tau) \sim \frac{1}{3 (\tau - \tau_0)} + \mbox{regular part}\;,
\end{equation}
where $\tau$ is a proper distance along $U_{reg}$ at $v< v_0$ and a
proper time at $v > v_0$, with $\tau = \tau_0$ corresponding to
$v=v_0$. We see that the Gibbons-Hawking integral~\eqref{eq:3}
diverges at the light-like point $\tau = \tau_0$, and there is no
apparent way to regularize it, cf.~\cite{Anderson:1986ww,
    Ishibashi:2002ac}.  

As a second try, we can leave the angles $S$ and $S'$ as they are
and simply ignore their Gibbons-Hawking contributions. But then arbitrary local
counter-terms $S_{S}$ and $S_{S'}$ depending on all RST fields at the
respective points can be added to the effective action. Importantly,
the point  $S'$ is visible to the distant observer. As a consequence,
its counter-term sets boundary conditions for future evolution: the
apparent initial velocity of $U_2(v)$ and quantum state of matter
going along the last ray $S'L$.

To summarize, the point $S'$ is a naked singularity of the complete
solution. It adds arbitrary parameters to the problem and therefore
limits our understanding of evaporating black holes.

%%%%%%%%%%%%%%%%%%%%%%%%%%%%%%%%%%
\subsection{Thunderpop}
\label{sec:thund}

In Sec.~\ref{sec:beyond-endpoint-1} we have found the thunderpop~--- an
outgoing $\delta$-flux carrying small negative energy~\eqref{eq:67}
along the last ray, cf.~\cite{Russo:1992ax}. It is harmless and can be
ignored. There is a belief in the literature~\cite{Strominger:1994tn, Ishibashi:2002ac},
however, that the last ray in causal one-loop models always turns into
an infinite-energy ``thunderbolt'' singularity. Let us show, why this
is not the case in our model.

\begin{figure}
  %\hrule

  \vspace{5mm}
  \hspace{2cm} (a) \hspace{7.2cm} (b)

  \vspace{-7mm}
  \centerline{
    \includegraphics{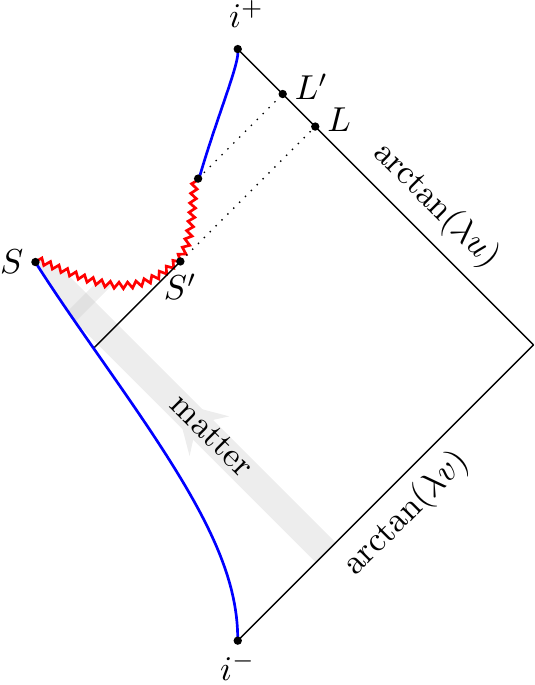} \hspace{20mm}
    \includegraphics{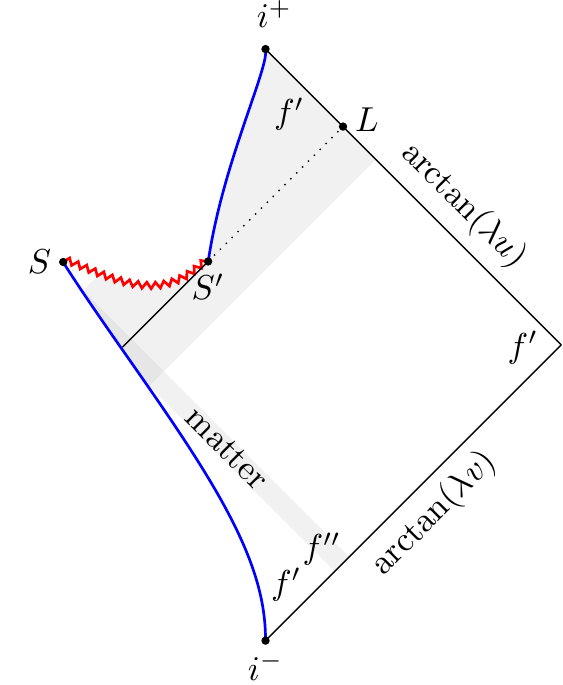}}
  
  % \hrule
  \caption{(a) Smearing the singularity $S'$. (b)~Global charge violation in the RST model.}
  \label{fig:non-conservation}
\end{figure}

For a start, we reproduce the argument
of~\cite{Strominger:1994tn}. Consider the quantum correlator of one RST
field, say, $f_N$. Initially, this field is in vacuum, 
$$
  \langle f_{N,\, \mathrm{in}}(\bar{v}_1) f_{N,\,
    \mathrm{in}}(\bar{v}_2) \rangle  = - \frac{1}{4\pi} \log|
  \bar{v}_2 - \bar{v}_1| + \mbox{const}\;.
$$
The outgoing two-point function then follows from the reflection
law~\eqref{eq:20},
\begin{align}
  \notag
  \langle f_{N,\, \mathrm{out}}(\bar{u}_1)
  f_{N,\, \mathrm{out}}(\bar{u}_2) \rangle_{reg} & \equiv \langle f_{N,\, \mathrm{out}}(\bar{u}_1)
  f_{N,\, \mathrm{out}}(\bar{u}_2) \rangle
  - \langle f_{N,\, \mathrm{out}}(\bar{u}_1)
  f_{N,\,  \mathrm{out}}(\bar{u}_2) \rangle_{vac}
  \\  \label{eq:58}
 & = -\frac{1}{4\pi} \left( \,
    \log|\bar{v}(\bar{u}_2) - \bar{v}(\bar{u}_1)| -
    \log|\bar{u}_2 - \bar{u}_1|\, \right) + \mbox{const}\;,
\end{align}
where we subtracted the vacuum correlator for regularity and introduced the boundary
$\bar{v}(\bar{u})$ in asymptotic coordinates.

If the boundary is smooth, the regularized correlator (\ref{eq:58}) is
finite in the limit ${\bar{u}_1\to\bar{u}_2}$ due to cancellation
between the first and second terms. However, 
$\bar{v}(\bar{u})$ has a jump at the last ray $\bar{u} = \bar{u}_e$,
see Fig.~\ref{fig:BH-exact-numerical}c. As a consequence, the first
term stays finite when $\bar{u}_{1}$ and $\bar{u}_2$ approach
$\bar{u}_e$ from different sides, and the second term is divergent. This
produces a strong non-integrable singularity in the energy
flux at $\bar{u} = \bar{u}_e$,
$$
\langle \bar{T}^{(N)}_{\bar{u}\bar{u}}(\bar{u}_e) \rangle_{reg} =
  \lim_{\bar{u}_2,\, \bar{u}_1 \to \bar{u}_e} \partial_{\bar{u}_2} 
  \partial_{\bar{u}_1}\langle f_{N,\, \mathrm{out}}(\bar{u}_1) 
  f_{N,\, \mathrm{out}}(\bar{u}_2) \rangle_{reg} \sim
  \lim_{\bar{u}_2,\, \bar{u}_1 \to \bar{u}_e}  \frac{1}{4\pi
    (\bar{u}_2 - \bar{u}_1)^2}\;,
  $$
where we used regularization by point separation. 

We have already argued, however, that the endpoint $S'$ is a naked
singularity in the one-loop model, and the quantum state of
the fields $f_i$ leaving this point is not under
control. Let us smear this point into a tiny timelike singularity
$\phi = \phi_0$, see Fig.~\ref{fig:non-conservation}a. Then $f_{N,\,
  \mathrm{out}}$ is not fixed at $\bar{u}_{L} < \bar{u} <
\bar{u}_{L'}$. One can therefore continue the regularized
correlator~\eqref{eq:58} into this region in an arbitrarily smooth
way, matching it together with its first two derivatives to
the correlators at $\bar{u}_{L}$ and $\bar{u}_{L'}$.  This will give
finite energy-momentum tensor inside the interval $LL'$. There is a restriction,
however: the jump of the primitive $\partial_u h$ between $\bar{u}_L$
and $\bar{u}_{L'}$ is fixed by the equations for the boundary, see
Eq.~\eqref{eq:66}. This gives the total energy $E_{\mathrm{th-pop}}
\approx - \lambda Q^2$ inside the ``quantum'' region $LL'$. Notably,
$E_{\mathrm{th-pop}}$ in vanishingly small in the quasi-stationary
limit \eqref{eq:4} that we consider.

%%%%%%%%%%%%%%%%%%%%%%%%%%%%%%%%%%%%%%%
\subsection{Absence of remnants}
\label{sec:absence-remnants}
Let us make the information paradox explicit in our model. To this end
we introduce geometric entropy $\Sigma_{reg}(\bar{u}_1,\,
\bar{u}_2)$ characterizing entanglement of the outgoing fields $\{
f_{i,\, \mathrm{out}}\}$ inside the interval $\bar{u}_1 < \bar{u} <
\bar{u}_2$ with anything outside
  it~\cite{Holzhey:1994we}. In Appendix~\ref{sec:ent-entropy-cft} we
evaluate the entropies of the Hawking quanta ${\Sigma^- \equiv
  \Sigma_{reg} (-\infty,\, \bar{u}_e - 0)}$ before the last ray and of
non-thermal radiation beyond the endpoint $\Sigma^+ \equiv
\Sigma_{reg} (\bar{u}_e + 0, +\infty)$, see
Fig.~\ref{fig:BH-exact-numerical}c. For $N = 48 \pi Q^2$ fields,
\begin{equation}
  \label{eq:42}
  \Sigma^{-} = -4\pi Q^2 \log
  \left.\frac{d\bar{v}}{d\bar{u}} \right|_{\bar{u} = \bar{u}_e - 0}\;,
  \qquad \qquad   \Sigma^{+} = -4\pi Q^2 \log
  \left.\frac{d\bar{v}}{d\bar{u}} \right|_{\bar{u} = \bar{u}_e + 0}\;,
\end{equation}
where $\bar{v}(\bar{u})$ is the boundary in the asymptotically flat 
coordinates~\eqref{eq:37}; at $\bar{u} < \bar{u}_e$ and $\bar{u} >
\bar{u}_e$ it represents $U(v)$ and $U_2(v)$, respectively.

If the state of the outgoing radiation was pure at $-\infty < \bar{u}
< +\infty$, the geometric entropies $\Sigma^-$ and $\Sigma^+$ would
coincide because their intervals are
complementary\footnote{For a time, we ignore the entropy of the
  thunderpop at $\bar{u} = \bar{u}_e$.}. This equality is automatic in
Eq.~(\ref{eq:42}) for any smooth $\bar{v}(\bar{u})$. Physically,
$\Sigma^+ = \Sigma^-$ would mean that the Hawking radiation is
entangled with the remnant~\cite{Almheiri:2013wka}~--- a state of
quantum fields beyond the endpoint. However our boundary has an
unavoidable jump at $\bar{u} = \bar{u}_e$, and we are going to
demonstrate that it makes the two entropies essentially different.

Start with $\Sigma^-$ for the Hawking radiation. To make
the estimate transparent, we assume that the incoming matter is well
localized, has large energy, and therefore collides with the boundary
after crossing the horizon in Fig.~\ref{fig:BH-exact-numerical}c. Then
$U(v)$ is approximately given by Eq.~\eqref{eq:57}, and we obtain,
\begin{equation}
  \label{eq:43}
  \Sigma^- \approx 4\pi Q^2 \log \left[-v_\times \partial_v
    U(v_\times)/u_e\right] \approx 4\pi Q^2 \log \frac{v_e}{v_\times}
    \approx \frac{4\pi}{\lambda} \, (E_{\mathrm{in}} - M_{cr})\;,
\end{equation}
where we introduced the crossing time $U(v_{\times}) = 0$, applied
Eq.~\eqref{eq:62}, and ignored the terms suppressed by $Q^2$. Expression
\eqref{eq:43} coincides with the thermal entropy of one-dimensional
gas with energy $E_{\mathrm{in}} - M_{cr}$ and temperature $T_{H} =  
\lambda / (2\pi)$. Thus, the entanglement entropy of the Hawking radiation
has the maximal possible value. It can be arbitrarily large at high
$E_{\mathrm{in}}$. This reproduces the standard result.

The ``remnant'' entropy  $\Sigma^+$ in Eq.~(\ref{eq:42}) depends on
the apparent initial velocity $\partial_v U_2(v_e)$ of the second
boundary, which is not fixed. In Sec.~\ref{sec:beyond-endpoint-1} we
demonstrated, however, that this velocity satisfies the
inequality~\eqref{eq:63}, or the spacetime would be singular beyond
the  endpoint. We thus obtain,
\begin{equation}
  \label{eq:45}
  \Sigma^+ = 4\pi Q^2 \log \left[-v_e \partial_v
    U_2(v_e)/u_e\right] < 4\pi Q^2 \log [- Q^2/(4\lambda^2 u_e v_e)]
  \sim O(Q^2)\;,
\end{equation}
where Eq.~\eqref{eq:62} was used in the last equality. Thus, 
non-thermal radiation beyond the endpoint is almost pure.
Parametric difference between Eqs.~(\ref{eq:43}) and (\ref{eq:45})
is the essence of the Hawking information paradox.

One can try to store the entanglement entropy into the thunderpop
which emanates from the endpoint singularity and can be in any
quantum state. For example, introducing $\tanh$-like smoothing of
$\bar{v}(\bar{u})$ at $\bar{u} = \bar{u}_e$, one automatically obtains $\Sigma^- =
\Sigma^+$ in Eq.~(\ref{eq:42}). The entropies remain equal even in the
limit when the smoothing region becomes small, $\delta \bar{u} \to
0$. However, the same smoothing introduces an infinite-energy
thunderbolt going along the last ray. Indeed, the
Schwarzian reflection law~\eqref{eq:23} gives,
\begin{equation}
\notag
  \int\limits_{\bar{u}_e - \delta \bar{u}}^{\bar{u}_e +\delta \bar{u}}
  d\bar{u}\; \bar{T}_{\bar{u}\bar{u}} \sim Q^2
  \int\limits_{\bar{u}_e - \delta \bar{u}}^{\bar{u}_e +\delta \bar{u}}
  d\bar{u} \, \left(\frac{\partial_{\bar{u}}^2 \bar{v}}{\partial_{\bar{u}}
    \bar{v}}\right)^2 \geq  \frac{Q^2}{2\delta \bar{u}}
  \left(\int_{\bar{u}_e - \delta\bar{u}}^{\bar{u}_e + \delta \bar{u}}
  \frac{\partial_{\bar{u}}^2 \bar{v}}{\partial_{\bar{u}} \bar{v}}\right)^2 =
  \left.\frac{(\Sigma^- - \Sigma^+)^2}{32 \pi^2 \,\delta \bar{u}\, Q^2}\right|_{unreg}\;,
\end{equation}
where we ignored all the terms regular at $\delta \bar{u} \to
  0$, used the Cauchy inequality, and expressed the result via
  the entropy mismatch between Eqs.~(\ref{eq:43}) and 
(\ref{eq:45}). We see that the energy of the ``thunderbolt'' diverges
at $\delta \bar{u}\to 0$. Moreover, ``Planckian'' thunderbolt with
$\delta \bar{u} \sim \lambda^{-1}$ has energy
$E_{\mathrm{in}}^2/(\lambda Q^2)$ which parametrically exceeds
the black hole mass. Thus, our small innocent thunderpop cannot
recover the information. To do that, radical large-distance modification
of the semiclassical geometry is needed.

  There is another curious property of the RST black holes which is
  apparent in our calculations. Recall that the black hole mass
  reaches a finite critical value $M \approx M_{cr} = 2\lambda
  \mathrm{e}^{-2\phi_0}$ at the endpoint of evaporation. Nevertheless,
  this critical black hole decays into an almost pure state with small
  entanglement entropy, see Eq.~\eqref{eq:45}. Thus, the
  thermodynamical entropy $S_{BH}$ of this object should be also
  small. Direct calculations of the black hole entropy~\cite{Fiola:1994ir,
    Myers:1994sg, Hayward:1994dw, Solodukhin:1995te} are consistent
  with this unusual property. They give expression,
\begin{equation} 
  \label{eq:35}
  S_{BH} = 4 \pi (\hat{\phi}_{hor} - \hat{\phi}_0)\;,
\end{equation}
which reaches zero at the endpoint when the value $\hat{\phi}_{hor}$
of the JT dilaton at the apparent horizon equals $\hat{\phi}_0$. However, while the
mass-dependent part of Eq.~(\ref{eq:35}) is fixed by the black hole
thermodynamics, the constant part is added
somewhat ad hoc, using additional physical considerations. Our results
independently confirm that $S_{BH}=0$ at $\phi_{hor} = \phi_0$. We are
going to further address this question in the forthcoming
publication~\cite{Fitkevich:2020tcj}.

%%%%%%%%%%%%%%%%%%%%%%%%%%%%%%%%%%%%%%%
\subsection{Non-conservation of a global charge}
\label{sec:glob-charge-nonc}

It has long been believed that quantum gravity does not tolerate any
conserved global charges. Indeed, by causality  black holes
evaporate into all sorts of particles regardless to what they were made 
of, violating all global quantum numbers.

The RST action~\eqref{eq:6},~\eqref{eq:3} has a global shift symmetry
$f\to f+ \mbox{const}$ and the respective conserved
current\footnote{Actually, $N$ currents~--- one per matter field.} $j^{\mu} =
\nabla^{\mu} f$. This gives the asymptotic conservation law, 
\begin{equation}
  \notag
  0 = \int d^2 x \,\sqrt{-g} \;\nabla_\mu j^\mu  = \int\limits_{\phi = \phi_0}
  d\tau \, 
  n^{\mu}\nabla_\mu f - \int d\bar{v} \, \partial_{\bar{v}} f_{\mathrm{in}}(\bar{v}) + \int
  d\bar{u} \, \partial_{\bar{u}} f_{\mathrm{out}}(\bar{u})\;,
\end{equation}
where we used the Gauss theorem with boundary terms coming from the
line $\phi = \phi_0$ and two light-like infinities. Recall that
$n^\mu$ and $\tau$ are the outer normal and proper time of the
boundary, whereas  $(\bar{u},\, \bar{v})$ represent asymptotically
flat coordinates. If the spacetime has simple topology like in
Fig.~\ref{fig:BH-exact-numerical}a, the matter fields satisfy Neumann
conditions at the boundary, and the conservation law
\begin{equation}
  \label{eq:47}
        {\cal Q} \equiv f_{\mathrm{in}}(\bar{v} = +\infty) -
  f_{\mathrm{in}}(\bar{v} = -\infty) = 
  f_{\mathrm{out}}(\bar{u} = +\infty)  - f_{\mathrm{out}}(\bar{u} = -\infty)
\end{equation}
holds. 
But once the black hole appears in Fig.~\ref{fig:BH-exact-numerical}c,
this law gets broken because ${n^\mu\nabla_\mu f \ne 0}$ at the spatial
sections of the line $\phi = \phi_0$.

Figure \ref{fig:non-conservation}b illustrates violation
of the charge~(\ref{eq:47}) by the evaporating black 
hole. In this figure the incoming wave packet $f_{\mathrm{in}}(\bar{v})$ approaches
$f'$ at
$\bar{v} \to \pm\infty$ taking the value $f''$ in between (grey
region). Its global charge is zero. But
reflection from the boundary gives $f_{\mathrm{out}} = 
f''$ at the endpoint $S'$ of black hole evaporation, and by
continuity\footnote{This does not contradict to the reflection law of
  $f$ which has an integration constant: ${f_{\mathrm{out}}(U(v)) =
  f_{\mathrm{in}}(v) + \mbox{const}}$, cf.\ Eq.~\eqref{eq:20}.}~--- in
the entire region beyond the last ray $S'L$. Then the global 
charge of $f_{\mathrm{out}}$ is ${\cal Q}_{\mathrm{out}} \equiv f'' - f' \ne 0$.

One can again try to pack the compensating charge inside the
thunderpop, to save the conservation law. But this requires energy,
like in the case with entropy. The energy of the thunderpop in
asymptotic coordinates is
\begin{equation}
  \notag
  E_{\mathrm{th-pop}}=\int\limits_{\bar{u}_e- \delta \bar{u}}^{\bar{u}_e + \delta
    \bar{u}} d\bar{u}\,(\partial_{\bar{u}}f_{out})^2 \geq
  \frac{1}{2\delta u} \left(\int_{\bar{u}_e-\delta
    \bar{u}}^{\bar{u}_e+ \delta \bar{u}} 
  d\bar{u}\,\partial_{\bar{u}}f_{out} \right)^2 =
  \frac{{\cal Q}^2_{\mathrm{th-pop}}}{2\delta \bar{u}}\;,
\end{equation}
where $\delta \bar{u}$ is the size of the uncontrollable region,
we used the Cauchy inequality and denoted ${\cal
    Q}_{\mathrm{th-pop}} = f_{\mathrm{out}}(\bar{u}_e + \delta
  \bar{u}) - f_{\mathrm{out}}(\bar{u}_e - \delta \bar{u})$. On the
other hand, the incoming energy is $E_{\mathrm{in}} \sim (f'' -
f')^2/\Delta \bar{v} = {\cal Q}_{\mathrm{out}}^2/\Delta \bar{v}$,
where $\Delta \bar{v}$ is a typical width of the incoming energy
flux. Thus, ${\cal Q}_{\mathrm{th-pop}} \ll {\cal
    Q}_{\mathrm{out}}$, if we want to keep $E_{\mathrm{th-pop}} <
E_{\mathrm{in}}$ and $\delta \bar{u} \ll \Delta \bar{v}$. The other
(unphysical) options would be to introduce a large naked
singularity with $\delta \bar{u}\sim \Delta \bar{v}$ or give the
thunderpop energy exceeding~$E_{\mathrm{in}}$.

We conclude that charge non-conservation in our model is
robust against quantum corrections at the last ray. 

%%%%%%%%%%%%%%%%%%%%%%%%%%%%%%%%%
\section{Discussion}
\label{sec:discussion}
In this paper we demonstrated local equivalence between the quantum
RST and flat-space JT models, and applied it to describe evaporating black holes. We
regularized the singularities of the RST fields with the reflective
boundary and derived one-loop effective action satisfying all
self-consistency requirements. Still, our semiclassical black hole solutions
are not satisfactory in three
respects. First, they violate the boundary conditions at the
spacelike line $\phi = \phi_0$ behind the horizon and therefore do
not extremize the effective action. This deprives the solutions from
their original role of saddle-point configurations for the path
integral and therefore makes all further physical interpretations
speculative. Second, the endpoint of evaporation corresponds to a
naked singularity with divergent action. Regularizing the action, one
introduces arbitrary counter-terms which determine the
subsequent evolution. Third 
and as a consequence of the first two, the final state of Hawking
radiation has large entanglement entropy indicating unitarity loss.

One can search for the root of the above problems in two
directions. The first suspect is the semiclassical method.  There are
many situations in semiclassical physics where the ``naive''
solutions do not exist: they are either singular~\cite{Affleck:1980mp}
or do not satisfy necessary boundary conditions~\cite{Levkov:2007yn},
just like the solutions in the RST model. In this case one can
apply the method of constrained instantons~\cite{Affleck:1980mp},
i.e.\ enforce correct behavior of the solutions with additional
constraint and then integrate over the constraint in the path
integral. This procedure was proposed for the black hole
evaporation~\cite{Bezrukov:2015ufa}, but has never been used
beyond the simplest thin shell models.

Also, one can make the semiclassical method
work by computing a different
quantity. For example, consider the unitarity relation,
\begin{equation}
  \label{eq:46}
  \mathrm{e}^{\int dk \, a_k^\dag b_k}  = \langle a | \hat{\cal S}^{\dag}
  \hat{\cal S}| b \rangle  = \int [dc^\dag  dc] \;\mathrm{e}^{-\int dk \,
  c_k^{\dag} c_k}\; \langle c | \hat{\cal
    S} | a \rangle^\dag \, \langle c |\hat{\cal S} | b\rangle\;,
\end{equation}
where $|a\rangle$, $|b\rangle$, and $|c\rangle$ are the coherent Fock
states of matter fields in flat spacetime with amplitudes $a_k$, $b_k$, and $c_k$, while
$\hat{\cal S}$ is the {${\cal S}$-matrix} of the RST model. At $a\ne 
b$ Eq.~(\ref{eq:46}) involves exponentially suppressed coherent
amplitudes of black hole formation and decay into a prescribed final state. The
processes of this kind are   described by complex semiclassical
solutions 
with distinct properties,  cf.~\cite{Rubakov:1996vz}
and~\cite{Berezin:1999nn, Parikh:1999mf,
  Bezrukov:2015ufa}. Using\footnote{The integral over $c_k$ and
  $c_k^\dag$ can be evaluated in  the saddle-point
  approximation.} the latter in Eq.~(\ref{eq:46}), one can directly
test unitarity of the RST ${\cal S}$-matrix.  

As a second possibility, one notes that the ``regularization''
boundary may ruin unitarity of the flat-space JT model. Indeed, the boundary
action~\eqref{eq:3} describes point particle stiffly coupled to the
dilaton field, with trajectory following the line $\phi = \phi_0$.
However, models with first-quantized relativistic particles are
generically non-unitary due to Klein
paradox~\cite{Klein:1929zz}: computation of transition
probabilities in these models gives unphysical results at
energies exceeding the threshold for particle-antiparticle
production. Our RST solutions demonstrate similar behavior. To remedy
this inconsistency, one can try to second-quantize the boundary, promoting
it to a quantum field. This may restore unitarity due to
boundary-antiboundary production.

%%%%%%%%%%%%%%%%%%%%%%%%%%%%%%%%
\paragraph{Acknowledgments.} We are grateful to Dmitry Eremeev for
  collaboration at early stages of this project. We also thank Sergey
  Sibiryakov, and Sergei Dubovsky for fruitful
  discussions. D.L.\ thanks Universit\'e libre de Bruxelles for
  hospitality. This work was supported by the grant RSF 16-12-10494.
%%%%%%%%%%%%%%%%%%%%%%%%%%%%%%%%
\appendix
%%%%%%%%%%%%%%%%%%%%%%%%%%%%%%%%
\section{Deriving the semiclassical equations}
\label{sec:cl-bulk-eq}
\subsection{Solution in the bulk}
\label{sec:derivation}
We obtain the semiclassical field equations by extremizing the
effective action~(\ref{eq:full-action}) with respect to the
background fields. In particular, the matter fields $f= f_1$ and
$\chi$ satisfy,
\begin{equation}
  \label{eq:19}
  \Box f=0\;, \qquad\qquad \Box\chi+QR=0\;,
\end{equation}
while variation with respect to $g^{\mu\nu}$ gives,
\begin{equation}
  \label{eq:24}
  2(2 \mathrm{e}^{-2\phi} + Q^2) \left(\nabla_\mu \nabla_\nu -
  g_{\mu\nu}\Box\right) \phi + 4 g_{\mu\nu}\mathrm{e}^{-2\phi} \left[
    (\nabla \phi)^2 - \lambda^2\right] = {\cal T}_{\mu\nu}^{(f)} +
  {\cal T}_{\mu\nu}^{(\chi)}\;.
\end{equation}
In the last equation we introduced the energy-momentum tensors,
\begin{align}
&{\cal T}^{(f)}_{\mu\nu}=\nabla_\mu f\nabla_\nu f-\frac12 g_{\mu\nu}(\nabla f)^2\;, 
\label{eq:en-mom_tensor_f}\\
&{\cal T}^{(\chi)}_{\mu\nu}=\nabla_\mu\chi\nabla_\nu\chi
-\frac12 g_{\mu\nu}(\nabla\chi)^2+2Q\left(\nabla_\mu\nabla_\nu - g_{\mu\nu}\square
\right)\chi\;, \label{eq:en-mom_tensor_chi}
\end{align}
which are conserved due to Eqs.~\eqref{eq:19}: $\nabla^\mu
{\cal T}_{\mu\nu}^{(f)} = \nabla^\mu {\cal T}_{\mu\nu}^{(\chi)} = 0$.
Semiclassical equation for $\phi$ can be written in the form
\begin{equation}
  \label{eq:25}
  \Box \phi + R/2 = 0
\end{equation}
using Eqs.~(\ref{eq:19}) and \eqref{eq:24}. 

Note that the RST vacuum~\eqref{eq:36} satisfies Eqs.~(\ref{eq:19}),
(\ref{eq:24}), (\ref{eq:25}) because in flat RST
coordinates~(\ref{eq:37}) it gives $\partial_\mu \partial_\nu \phi = 0$,
$(\partial_\mu \phi)^2 = \lambda^2$, $R = 0$, and $\chi=0$ due to
  Eq.~\eqref{eq:f_chi_solution}.

In the conformal gauge~(\ref{eq:7}) Eqs.~(\ref{eq:19}) and
(\ref{eq:25}) simplify, 
\begin{equation}
  \label{eq:21}
  \partial_u \partial_v f = 0\;, \qquad \qquad
  \partial_u \partial_v (\chi - 2Q \rho) = 0\;, \qquad \qquad 
  \partial_u \partial_v (\phi - \rho) =0\;,
\end{equation}
where we substituted $R= 8\mathrm{e}^{-2\rho}\, \partial_u \partial_v
\rho$. Solutions of the first two equations are given in
Eq.~(\ref{eq:f_chi_solution}), where the in- and outgoing wave packets
$f_{\mathrm{in}}(v)$, $\chi_{\mathrm{in}}(v)$ and
$f_{\mathrm{out}}(u)$, $\chi_{\mathrm{out}}(u)$ are arbitrary. The
last equation together with the residual reparametrization
invariance $u \to \tilde{u}(u)$, $v\to \tilde{v}(v)$ allows us to impose the
``Kruskal'' gauge $\rho=\phi$ in Eq.~\eqref{eq:34}. 

Substituting the matter fields~(\ref{eq:f_chi_solution}) into
Eqs.~(\ref{eq:en-mom_tensor_f}), (\ref{eq:en-mom_tensor_chi}), 
one computes their energy-momentum tensors,
\begin{equation}
  \label{eq:26}
  {\cal T}_{uu}^{(f)} + {\cal T}_{uu}^{(\chi)} = 4Q^2\left[\partial_u^2 \rho -
    (\partial_u \rho)^2\right] + T_{uu}\;,\qquad 
  {\cal T}_{uv}^{(f)} + {\cal T}_{uv}^{(\chi)} = -4Q^2 \partial_u \partial_v \rho\;,
\end{equation}
where expression for the $(vv)$ component can be obtained by replacing
$u\to v$. In Eq.~(\ref{eq:26}) we separated the 
contributions $T_{uu}(u)$ and  $T_{vv}(v)$ of the incoming and outgoing matter wave
packets defined in Eq.~(\ref{eq:27}). By themselves, $T_{vv}$ and
$T_{uu}$ are not tensors. However, they coincide with the energy
fluxes at infinity in the asymptotically flat  ($\rho=0$)
coordinates~$\bar{u}$ and~$\bar{v}$.

Using Eqs.~(\ref{eq:26}) in the ``Kruskal'' gauge $\rho = \phi$, we
rewrite Eq.~(\ref{eq:24}) as 
\begin{equation}
  \label{eq:28}
  \partial_u^2 \hat{\phi} = - T_{uu}/2 \;, \qquad \partial_v^2
  \hat{\phi} = - T_{vv}/2\;, \qquad \partial_u \partial_v
  \hat{\phi} = - \lambda^2\;.
\end{equation}
Recall that $\hat{\phi} = \mathrm{e}^{-2\phi} + Q^2 \phi$ is the JT
field introduced in Eq.~(\ref{eq:5}). General solution of
Eq.~(\ref{eq:28}) is
\begin{equation}
  \label{eq:38}
  \hat{\phi} = -\lambda^2 uv + c_1 + c_2 u + c_3 v + g(v) + h(u)\;,
\end{equation}
where the primitives $g$ and $h$ are given by Eqs.~(\ref{eq:30}),
(\ref{eq:35s}).

In the main text we consider solutions starting from the RST vacuum
\eqref{eq:36} in the asymptotic past. This means that the spacetime is flat,
$R=0$, and $\phi = -\lambda \bar{x}$ in the
beginning, see Eq.~\eqref{eq:37}. On the other hand, the coordinates
$u$ and $v$ cover semi-infinite intervals in Fig.~\ref{fig:intro}b. We
can shift them to the domains $-\infty < u < 0$ and $0<v<
+\infty$. Then the past time  infinity $i^-$ is reached at 
$u\to -\infty$, $v\to 0$, and the dilaton field
$\hat{\phi} \to -\lambda^2 uv - Q^2 \log(-\lambda^2 uv)/2 + c_1 + c_2 u$
should be a finite function of $\bar{x}$ in this limit, see Eq.~(\ref{eq:38}). Thus,
$c_2=0$ and $uv$ is finite 
at $i^-$. Moreover, we compute the curvature plugging
the past asymptotics of $\hat{\phi}$ into Eq.~(\ref{eq:25}) and find, 
\begin{equation}
  R \to - \mathrm{e}^{-2\phi}(\mathrm{e}^{-2\phi} + \lambda^2 uv) \; \frac{ (Q^4 +
    4\lambda^2 uv \mathrm{e}^{-2\phi})}{uv (Q^2/2 -
    \mathrm{e}^{-2\phi})^3}
  \qquad \mbox{as} \qquad u\to -\infty\;.
\end{equation}
Thus, in the regime~\eqref{eq:4} the spacetime is flat in the past
only if $\mathrm{e}^{-2\phi} = -\lambda^2 uv$. This gives $c_1 =0$ in
Eq.~(\ref{eq:38}). Note finally that the future time infinity $i^+$
is reached at $v\to +\infty$, $u\to 0$ and finite $\hat{\phi}$. This
is possible only if $c_3=0$ in Eq.~\eqref{eq:38}. At $c_1 = c_2 = c_3
=0$ the solution~(\ref{eq:38}) reduces to Eq.~(\ref{eq:29}) from the
main text.

%%%%%%%%%%%%%%%%%%%%%%%%%%%%%%%%%%%%%%%%%%%
\subsection{Reflection laws}
\label{sec:boundary-eq}
We obtain boundary conditions by varying the effective action
\eqref{eq:full-action} with respect to the boundary
values of all fields. Due to reparametrization invariance we can
consider only particular variations preserving the coordinate position of the
boundary: $\delta \phi = 0$ there.
Then~\cite{Poisson, Fitkevich:2017izc, Eremeev}
\begin{multline}
  \label{eq:52}
  \delta S=\int d\tau\left\{h^{\mu\nu} \delta h_{\mu\nu}
  \left[\nabla_n(\mathrm{e}^{-2\phi} - Q^2 \phi + Q\chi ) - \mu/2 \right]
  \right. \\ \left. - \delta \chi (\nabla_n \chi - 2KQ) - \delta f \nabla_n f
    \right\}\,,
\end{multline}
where we left only the boundary terms, introduced the outer normal derivative
${\nabla_n \equiv n^\mu \nabla_{\mu}}$ and the induced metric 
$h_{\mu\nu}\equiv g_{\mu\nu}-n_\mu n_\nu$. As before, we keep only one
matter field $f=f_1$. Then the semiclassical
boundary conditions are,
\begin{equation}
  \label{eq:22}
  \nabla_n \chi = 2QK\;, \qquad 
  \nabla_n f = 0\;,\qquad
  \left(2\mathrm{e}^{-2\phi_0} + Q^2\right) (\nabla_n \phi - \lambda)
  = 2Q^2 K\;, 
\end{equation}
where the value of $\mu$ was taken from
Sec.~\ref{sec:one-loop-corrections}.

Note that the flat RST vacuum~\eqref{eq:36}, \eqref{eq:37} with the static
boundary~(\ref{eq:static-boundary}) satisfies Eqs.~(\ref{eq:22}): in this
case $K=0$, $\chi=0$ and $\nabla_n \phi = \lambda$.  

\begin{sloppy}
  
In ``Kruskal'' coordinates $(u,\, v)$ with $\rho = \phi$ the boundary is
described by the  function $u = U(v)$. The outer normal has 
components
\begin{equation}
  \label{eq:53}
  \{n^u,\, n^v\} = \mathrm{e}^{-\phi}\{(\partial_v U)^{1/2}, \,
  -(\partial_v U)^{-1/2} \}\;,
\end{equation}
  where $\partial_v U>0$ for
the timelike boundary. Substituting the extrinsic curvature ${K
= \nabla_\mu n^\mu = n^\nu \partial_\nu \phi - \mathrm{e}^{-\phi}
\partial_v (\partial_v U)^{-1/2}}$ into the first two of
Eqs.~(\ref{eq:22}), we obtain the boundary conditions
$$
\partial_v U\,  \partial_u f_{\mathrm{out}} = \partial_v f_{\mathrm{in}}\;,
 \qquad
  \partial_v U \partial_u \chi_{\mathrm{out}} = \partial_v \chi_{\mathrm{in}} +
 Q \,\partial_v \log (\partial_vU) \qquad \mbox{at} \qquad u = U(v)\;.
$$
 Solutions of these equations are the reflection laws of
matter fields~(\ref{eq:20}) and of their energy fluxes~\eqref{eq:23}.

\end{sloppy}

The latter reflection laws together with the bulk
constraint give equation for $U(v)$. Indeed, full
derivative of $\hat{\phi}(U(v),\, v)$ along the boundary is
zero, hence, 
\begin{equation}
  \label{eq:32}
  \partial_u h - \lambda^2 v = -(\partial_v g - \lambda^2
  U)/\partial_v U \qquad \mbox{at}\qquad u = U(v)\;,
\end{equation}
where Eq.~\eqref{eq:29} was used. We take derivative
  of Eq.~\eqref{eq:32} along the boundary and then express $\partial_u^2 
  h$ from Eqs.~\eqref{eq:35s}, \eqref{eq:23}. We find,
\begin{equation}
  \label{eq:33}
  \partial_v\left(\frac{\partial_vg-\lambda^2U}{\sqrt{\partial_vU}}\right)=
  \frac{Q^2}{2} \,
  \partial_v\left(\frac{\partial_v^2 U}{(\partial_v U)^{3/2}}\right)\;.
\end{equation}
Integrating this equation, one arrives at Eq.~(\ref{eq:boundary_ode_wz})
with arbitrary constant in front of the right-hand side. 

The last boundary condition in Eqs.~(\ref{eq:22}) fixes the value of the
multiplicative constant in Eq.~(\ref{eq:boundary_ode_wz}). 
Indeed, we have already shown that the RST
vacuum~\eqref{eq:36},~(\ref{eq:static-boundary}) satisfies the entire set
of semiclassical equations and boundary conditions. On the other hand,
$U(v)$ in Eq.~(\ref{eq:static-boundary})  agrees with
Eq.~(\ref{eq:boundary_ode_wz}) only if the constant in
that equation equals $\mathrm{e}^{2\phi_0} q^2/\lambda^2$.  One can
explicitly check that once this value  is fixed, the third
of Eqs.~(\ref{eq:22}) is equivalent to 
Eq.~(\ref{eq:boundary_ode_wz}).

%%%%%%%%%%%%%%%%%%%%%%%%%%%%%%%%
\subsection{Energy conservation}
\label{sec:energy-conservation}
We introduce the energies $E_{\mathrm{in}}$ and $E_{\mathrm{out}}$ of
the incoming and outgoing matter by recalling that the fluxes
$\bar{T}_{\bar{u}\bar{u}}$ and $\bar{T}_{\bar{v}\bar{v}}$ coincide
with the respective components of the energy-momentum tensor ${\cal
  T}_{\mu\nu}$ in the asymptotically flat coordinates $(\bar{u},\,
\bar{v})$ with $\rho = 0$, see
Eq.~\eqref{eq:26}. Thus,
\begin{equation}
  \label{eq:39}
  E_{\mathrm{in}} = \int_{-\infty}^{+\infty} d\bar{v} \;
  \bar{T}_{\bar{v}\bar{v}} (\bar{v}) \;, \qquad \qquad
  E_{\mathrm{out}} = \int_{-\infty}^{+\infty} d\bar{u} \;
  \bar{T}_{\bar{u}\bar{u}} (\bar{u}) \;.
\end{equation}
Transforming to the ``Kruskal'' frame by
Eqs.~\eqref{eq:v-conformal-transformation}, \eqref{eq:37}, one gets, 
\begin{equation}
  \label{eq:41}
  E_{\mathrm{in}}=-2\lambda\int_0^{+\infty}dv\,v\left(\partial_v^2g -
  \frac{Q^2}{2v^2}\right)\;, 
  \;\;\;\;
  E_{\mathrm{out}}=2\lambda\int_{-\infty}^0 du\,u\left(\partial_u^2h -
  \frac{Q^2}{2u^2}\right)\;, 
\end{equation}
where $g(v)$ and $h(u)$ are the second primitives of $T_{vv}$ and
$T_{uu}$ in Eqs.~\eqref{eq:30}, \eqref{eq:35s}.

Let us demonstrate that the energy is conserved,
$E_{\mathrm{in}} = E_{\mathrm{out}}$, for the semiclassical solutions
starting from the RST vacuum in the past and arriving to it in the
asymptotic future. A black hole in the intermediate state, if it
evaporates completely, does not affect this conservation law. 

We introduce the integration limits $v_1 \to 0$, $v_2 \to
+\infty$, $u_1 \to -\infty$, and $u_2 \to 0$ in
Eq.~(\ref{eq:41}). Since the trajectory of the boundary approaches
the vacuum solution~\eqref{eq:static-boundary} at $v\to 0$ and $v\to
+\infty$, we choose $u_1 = U(v_1)$ and $u_2 = U(v_2)$. Performing the
integrals in Eq.~(\ref{eq:41}), one obtains~\cite{Eremeev},
$$
\frac{E_{\mathrm{in}}-E_{\mathrm{out}}}{2\lambda}= \left\{g(v) 
+
h(U(v))+\frac{Q^2}{2}\log\,(-\lambda^2vU(v))-v\partial_vg-U\partial_uh\right\}\Bigg|_{v_1}^{v_2}=0\;,
$$
where in the last equality we used Eqs.~\eqref{eq:static-boundary},
\eqref{eq:29}, recalled that $\phi = \phi_0$ along the boundary, and
evaluated the limits $v_1 \to 0$, $v_2 \to +\infty$. This proves energy
conservation.

%%%%%%%%%%%%%%%%%%%%%%%%%%%%%%%%%%%%%%%%%%%
\subsection{Equations for solvable deformation}
\label{sec:solvable-deformation}
Consider the model~\cite{Eremeev} with Dirichlet boundary
condition~\eqref{eq:51} 
for the quantum field~$\chi$. Boundary variation of the metric in Eq.~\eqref{eq:52} 
gives, 
$$
n^\mu\nabla_\mu\left(e^{-2\phi}-Q^2\phi+Q\chi\right)=
 - \lambda\left(2e^{-2\phi_0}+Q^2\right)\;.
 $$
Using the outer normal~\eqref{eq:53}, dilaton~\eqref{eq:29},
Eq.~\eqref{eq:32}, and reflection law for $\chi$ in the gauge $\rho =
\phi$, we obtain equation,
\begin{equation}\
  \label{eq:boundary_sd} 
 \partial_v U=e^{2\phi_0} \, 
\frac{q^2}{\lambda^2}
\left(\partial_vg+Q\partial_v \chi_{\mathrm{in}}-\lambda^2U\right)^2\;.
\end{equation}
We take the incoming quantum field in vacuum, ${\chi_{\mathrm{in}} = Q\, \log 
(\lambda v)}$, and arrive to Eq.~\eqref{eq:boundary_ode_sd} from the main
text.

%%%%%%%%%%%%%%%%%%%%%%%%%%%%%%%%
\section{Entanglement entropy}
\label{sec:ent-entropy-cft}
In this Appendix we review the geometric entropy $\Sigma(\bar{v}_1,\,
\bar{v}_2)$ measuring entanglement of the quantum field inside the
interval $[\bar{v}_1,\, \bar{v}_2]$ with the rest of the world. In the
particular case of one massless scalar field  in the vacuum state
this entropy equals~\cite{Holzhey:1994we},
\begin{equation}\label{eq:geom_entropy1}
\Sigma_{vac}(\bar{v}_1,\, \bar{v}_2)=\frac1{12}\; \log\left[\frac{(\bar{v}_2
    -   \bar{v}_1)^2}{\delta \bar{v}_1\delta \bar{v}_2}\right]\;, 
\end{equation}
where we restricted attention to the left-moving sector and introduced 
UV cutoffs $\delta \bar{v}_1$, $\delta \bar{v}_2$ at the borders of the
interval. In quantum theory~(\ref{eq:geom_entropy1}) is a
divergent quantity which should be renormalized. In thermodynamics,
$\delta\bar{v}_1$ and $\delta \bar{v}_2$ are the parameters of 
coarse-graining.

Importantly, the entanglement entropy is conformally
invariant~\cite{Holzhey:1994we}. Indeed, general conformal
transformation is local. It may distort the state of the quantum field but
never changes the degrees of freedom inside the interval. Selecting
one RST field~--- say, $f_N$~--- one recalls that it is initially in
vacuum and its reflection from the boundary is a conformal
transformation~\eqref{eq:20}. Thus, the outgoing entropy of this field
is, 
\begin{equation}\label{eq:geom_entropy2}
\Sigma(\bar{u}_1, \,
\bar{u}_2)=\Sigma_{vac}(\bar{v}(\bar{u}_1),\,
\bar{v}(\bar{u}_2))=\frac1{12}\log\,\left[
\frac{(\bar{v}(\bar{u}_2) -
  \bar{v}(\bar{u}_1))^2}{\delta \bar{u}_1 \delta
  \bar{u}_2\, \bar{v}'(\bar{u}_1)\,
  \bar{v}'(\bar{u}_2) \,}\right]\;,
\end{equation}
where we introduced the trajectory $\bar{v} = \bar{v}(\bar{u})$ of the
boundary in asymptotic coordinates and related the cutoffs $\delta
\bar{v}(\bar{u}_i)$ to $\delta \bar{u}_i$; primes are the
$\bar{u}$-derivatives. We stress that the outgoing field $f_{N,\,
  \mathrm{out}}$ with entropy~(\ref{eq:geom_entropy2}) is not in
vacuum.

Finally, we regularize the entropy~(\ref{eq:geom_entropy2}) by
subtracting its vacuum value, 
\begin{equation}\label{eq:ren_entropy}
\Sigma_{reg}(\bar{u}_1,\, \bar{u}_2)=
\Sigma(\bar{u}_1,\, \bar{u}_2) -
\Sigma_{vac}(\bar{u}_1,\,
\bar{u}_2)=\frac1{12}\log\left[\frac{(\bar{v}(\bar{u}_2)-\bar{v}(\bar{u}_1))^2
}{(\bar{u}_2-\bar{u}_1)^2\, \bar{v}'(\bar{u}_2)\, \bar{v}'(\bar{u}_1)}\right]\;.  
\end{equation}
This quantity measures entanglement of the outgoing field within the
interval $[\bar{u}_1,\;\bar{u}_2]$ relative to the vacuum. In the main
text we use half-infinite intervals $(-\infty,\, \bar{u})$
and $(\bar{u},\, +\infty)$. Sending $\bar{u}_2 \to \pm \infty$  and
recalling that in this limit ${\bar{v}(\bar{u}_2) \to \bar{u}_{2} -
  2\phi_0 / \lambda}$, we obtain
\begin{equation}
    \Sigma_{reg}(-\infty,\, \bar{u}) = \Sigma_{reg}(\bar{u},\, +\infty)
    = - \frac{1}{12}\log \bar{v}'(\bar{u})\;.
\end{equation}
Multiplication by the number of fields $N = 48\pi Q^2$ gives
Eq.~\eqref{eq:42}.

%%%%%%%%%%%%%%%%%%%%%%%%%%%%%%%%%%%%%%%%%

\end{document}